\documentclass[preprint,12pt]{elsarticle}

\usepackage{amssymb}
\rmfamily{\fontsize{12pt}{\baselineskip}\selectfont}

\usepackage{graphicx}
\usepackage{amsfonts, amsmath, amssymb, amsbsy, amsthm}
\usepackage{bm}
\usepackage{color}
\usepackage{xcolor}
\usepackage{booktabs}
\usepackage{tikz}
\usetikzlibrary{shapes.geometric, arrows}
\tikzstyle{block} = [rectangle, draw, fill=gray!20, text width=7em, text centered, rounded corners, minimum height=2em, font=\small]
\tikzstyle{line} = [draw, -latex']
\tikzstyle{dashedline} = [draw, -latex', dashed]  
\usepackage{hyperref}
\usepackage{mathrsfs}
\usepackage{longtable}
\usepackage[top=2.5cm,bottom=2.0cm,left=2.0cm,right=2.0cm]{geometry}
\usepackage{epstopdf}
\usepackage{stmaryrd}
\usepackage[normalem]{ulem}
\usepackage{cancel}
\usepackage{scalerel}
\usepackage{enumitem}
\usepackage{algorithm}
\usepackage{subfigure}
\usepackage{algorithm, algorithmicx}
\usepackage[noend]{algpseudocode}
\usepackage{lineno}
\usepackage{enumitem}
\usepackage{diagbox}
\usepackage{multirow}
\usepackage{siunitx}
\usepackage{booktabs}
\renewcommand{\algorithmiccomment}[1]{\bgroup\hfill//~#1\egroup}

\def\ml{\textsf{MACE-MP-0b}}
\def\mlo{\textsf{MACE-MP-0}}

\begin{document}

\begin{frontmatter}

\title{A Study on the Fine-Tuning Performance of 
Universal Machine-Learned Interatomic Potentials (U-MLIPs)}

\author[sciaiaddress]{Xiaoqing Liu}
\ead{liuxiaoqing@sciai.com.cn}

\author[sciaiaddress]{Kehan Zeng}

\author[nusaddress]{Yangshuai Wang}

\author[sciaiaddress,INSaddress]{Teng Zhao}


\address[sciaiaddress]{Shanghai Jiao Tong University-Chongqing Institute of Artificial Intelligence, Chongqing 401329, China.}

\address[ubcaddress]{Department of Mathematics, University of British Columbia, 1984 Mathematics Road, Vancouver, Canada.}

\address[nusaddress]{Department of Mathematics, National University of Singapore, Singapore.}

\address[INSaddress]{Institute of Natural Sciences, MOE-LSC, and Shanghai National Center for Applied Mathematics, Shanghai Jiao Tong University, Shanghai 200240, China.}

\begin{abstract}
Universal machine-learned interatomic potentials (U-MLIPs) have demonstrated effectiveness across diverse atomistic systems but often require fine-tuning for task-specific accuracy. We investigate the fine-tuning of two MACE-based foundation models, \mlo~and its variant~\ml, and identify key insights. Fine-tuning on task-specific datasets enhances accuracy and, in some cases, outperforms models trained from scratch. Additionally, fine-tuned models benefit from faster convergence due to the strong initial predictions provided by the foundation model. The success of fine-tuning also depends on careful dataset selection, which can be optimized through filtering or active learning. We further discuss practical strategies for achieving better fine-tuning foundation models in atomistic simulations and explore future directions for their development and applications.
\end{abstract}





\end{frontmatter}

\section{INTRODUCTION}
\label{sec:intro}

Machine learning (ML) has transformed research paradigm of atomistic materials science, particularly in molecular dynamics (MD) simulations~\cite{behler2016perspective, vamathevan2019applications, raccuglia2016machine}. While Density Functional Theory (DFT)~\cite{sholl2022density, koch2015chemist} provides high accuracy, its computational cost is prohibitive for large systems and long-time simulations. Traditional empirical force fields~\cite{ponder2003force, altona2005empirical, lee2000second} are computationally efficient but often fail to capture complex interatomic interactions. This limitation highlights the need for ML-driven models that achieve DFT-level accuracy with significantly lower computational cost.

Machine-learned interatomic potentials (MLIPs) bridge the gap between DFT and traditional force fields~\cite{behler2007generalized, bartok2010gaussian, witt2023acepotentials, batatia2022mace, shapeev2016moment}. The MACE architecture~\cite{batatia2022mace}, built on atomic cluster expansion (ACE)~\cite{DrautzACE, ACECompleteness}, improves efficiency through tensor decomposition~\cite{luo2024enabling, darby2023tensor} and higher-order equivariant message-passing~\cite{gilmer2017neural, gilmer2020message}. It features high body order equivariant representations, requiring only two message-passing layers, and minimal nonlinearity, classifying it as a graph tensor network. Recent advances in large language models~\cite{thirunavukarasu2023large, kasneci2023chatgpt} have facilitated pre-trained MLIPs trained on extensive materials datasets~\cite{chanussot2021open, bowman2022md17, jain2013commentary}, leading to foundation models~\cite{bommasani2021opportunities} that simplify adoption and provide strong baselines for fine-tuning. \mlo~\cite{batatia2023foundation}, a MACE-based pre-trained model, is trained primarily on DFT relaxation trajectories, while other architectures and datasets have yielded alternative models with distinct capabilities~\cite{deng2023chgnet, merchant2023scaling, zhang2023dpa, choudhary2023unified, chen2022universal}.

While \mlo~has shown robustness in MD simulations~\cite{batatia2023foundation}, its accuracy in predicting atomic interactions and energy landscapes remains limited, particularly for mechanical properties like elastic constants and stacking fault energies in elemental alloys~\cite{li2024extendable}. This highlights the need for fine-tuning to improve predictive accuracy. Fine-tuning adapts pre-trained models to specific systems, refining their parameters to better align with high-fidelity computational or experimental data. While research on fine-tuning atomistic foundation models is expanding, comprehensive studies on key observations, comparisons with training from scratch, and potential applications remain limited. Recent works have started addressing these aspects, and interested readers may refer to~\cite{focassio2024performance, deng2024overcoming, alavi2024towards, yu2024systematic, pyzer2025foundation, shuang2025universal, du2025universal, lee2025accelerating, niblett2024transferability, casillas2024evaluating}.

In this work, we evaluate the fine-tuning performance of two MACE-based foundation models, \mlo~and its variant~\ml. Our main observations are: (1) While these models show good zero-shot accuracy and stability in MD simulations, their performance on specific tasks can be limited. Fine-tuning on task-specific datasets improves accuracy and can sometimes match or slightly surpass models trained from scratch. (2) In some cases, fine-tuning converges faster than training from scratch, likely because the foundation model provides a reasonable starting point. (3) Dataset selection is important for fine-tuning. Filtering from candidate datasets or using active learning strategies can be beneficial, consistent with recent findings~\cite{wang2024impact} that emphasize the value of specialized datasets over increased model complexity. This work presents practical observations on fine-tuning foundation models for atomistic simulations and explores potential directions for future studies, including model distillation for improved efficiency.

It is worth noting that the fine-tuning of U-MLIPs can be efficiently carried out using the Random-Batch Molecular Dynamics (RBMD) package, which enables large-scale particle simulations at the nano- and microscale. In contrast to conventional molecular dynamics frameworks, RBMD leverages random batch algorithms~\cite{RBM_Jin, RBE, RBE_2, IRBE} to efficiently handle nonbonded interactions, supporting simulations of up to 10 million particles on a single GPU with a CPU core~\cite{gao2024rbmd}. Moving forward, we aim to integrate the RBMD platform with MACE-based MLIPs for more practical and large-scale simulations. To support this integration, an open-access platform is available at \url{https://www.randbatch.com/rbmd}, where further examples and implementation scripts for realistic simulation scenarios will be released in the future.

\section{METHODS}
\label{sec:methods}

\subsection{The MACE Architecture}
\label{sec:sub:mace}

MLIPs map atomic positions, chemical elements or any additional information such as partial charges to the potential energy surface for a given atomic system. The MACE model~\cite{batatia2022mace} extends the ACE framework~\cite{witt2023acepotentials, DrautzACE, ACECompleteness, ho2024atomic}, originally developed for MLIPs but now applied beyond atomistic modeling~\cite{ACEHam, wang2024theoretical, torabi2024surrogate, wang2025many}. MACE is an equivariant message-passing graph tensor network~\cite{vignac2020building, maron2018invariant} that encodes many-body atomic geometry at each layer. It constructs many-body messages using a linear combination of tensor product bases, formed from two-body permutation-invariant polynomials expanded in a spherical basis~\cite{DrautzACE, ACECompleteness}. Equivariance is preserved through tensor contraction of irreducible representations with generalized Clebsch-Gordan coefficients~\cite{luo2024enabling, darby2023tensor}. The final output on each atom represents corresponding contributions to the total potential energy. For a detailed description of the architecture, we refer to~\cite{batatia2022mace}.

Compared to other message-passing neural network potentials, MACE introduces two key innovations: (1) it incorporates high body order equivariant features (4-body in this case) in each layer, making two message-passing layers sufficient; (2) it is only mildly nonlinear, with nonlinearity restricted to the radial basis and final readout layer, classifying it as a graph tensor network. Its computational cost is comparable to other graph neural network potentials, enabling simulations of around a thousand atoms for nanoseconds per day on a GPU~\cite{batatia2022mace}.

\subsection{The Foundation Models}
\label{sec:sub:pre-trained}

While the MACE architecture is known for its accuracy, its foundation model \mlo, trained exclusively on the Materials Project dataset~\cite{jain2013commentary}, has demonstrated strong performance across various applications, particularly in enabling stable MD simulations across diverse chemical environments~\cite{batatia2023foundation}. \mlo~is available in three variants, distinguished by maximal message equivariance ($L=0,1,2$ for small, medium, and large models, respectively). This work employs the medium model ($L=1$) due to its emphasis on stress training and its favorable balance between computational cost and accuracy. 

A recent update, \ml, introduces some promotions, including improved pairwise repulsion and correct handling of isolated atoms. During the preparation of this manuscript, several additional foundation model versions were developed, focusing on enhancing stability under high pressure (\textsf{MACE-MP-0b2}), resolving phonon-related issues (\textsf{MACE-MP-0b3}), and incorporating a broader training dataset (\textsf{MACE-MPA-0} and \textsf{MACE-OMAT-0}). Readers interested in these models can find more details in the GitHub repository (\url{https://github.com/ACEsuit/mace-mp/}).

While these foundation models generally perform well, stable MD simulations alone do not guarantee quantitatively accurate physical properties. Fine-tuning remains essential for adapting the models to specific tasks. In this work, we focus on \mlo~and its first variant, \ml, as the fine-tuning strategies discussed here are expected to generalize to other variants.

\subsection{Multi-head Fine-tuning}
\label{sec:sub:ft}

Fine-tuning pre-trained models is widely used to enhance accuracy and generalization across various fields, including image analysis~\cite{tajbakhsh2016convolutional, jordan2015machine} and large language models~\cite{zhang2022fine, devlin2019bert, lee2020biobert}. In molecular modeling, fine-tuning follows a ``predictor-corrector" scheme. The foundation model, trained on diverse configurations, serves as an initial predictor, providing a universal force field, which is robust under reasonable MD setups. Fine-tuning then refines this model using a targeted dataset that captures system-specific interactions, improving accuracy in dynamic simulations. 

The choice of fine-tuning dataset is crucial. A common approach is active learning, where MD simulations generate target configurations for DFT calculations, iteratively expanding the dataset until satisfactory accuracy is reached. It is noted that, implementing a full active learning strategy requires robust uncertainty estimation, which is beyond the scope of this work. Instead, we consider three practical dataset selection methods summarized in Table~\ref{tab:config_methods}. FT-1 directly uses open-source configurations (Sections~\ref{sec:sub:metals}, \ref{sec:sub:disloc}, \ref{sec:sub:LiCl}). FT-2 applies manual filtering (Section~\ref{sec:sub:Si}). FT-3 uses uncertainty-based filtering on a self-constructed dataset (Section~\ref{sec:sub:HEA}), following~\cite{hyperactive2022}, with uncertainty estimated via Bayesian linear regression of an auxiliary ACE model~\cite{witt2023acepotentials, DrautzACE}. 

\begin{table}[h!]
\centering
\begin{tabular}{l|p{10cm}|l}
\toprule
\textbf{Name} & \textbf{Description} & \textbf{Application} \\ 
\midrule
FT-1 & Configurations are directly from open-source datasets without additional filtering. & Sections~\ref{sec:sub:metals}, \ref{sec:sub:disloc}, \ref{sec:sub:LiCl} \\ 
\midrule
FT-2 & Configurations selected from open-source datasets and refined through manual filtering. & Section~\ref{sec:sub:Si} \\ 
\midrule
FT-3 & Configurations selected by applying filtering with uncertainties to a self-constructed candidate dataset. & Sections~\ref{sec:sub:HEA} \\ 
\bottomrule
\end{tabular}
\caption{Summary of configuration selection methods for fine-tuning and their corresponding applications.}
\label{tab:config_methods}
\end{table}

Multi-head fine-tuning extends transfer learning by attaching multiple task-specific heads to a shared pre-trained backbone~\cite{voita2019analyzing, kim2024hydra}. This approach allows the model to handle multiple related tasks within a single framework, leveraging the general feature representations of the backbone while adapting each head to task-specific requirements. An implementation of multi-head fine-tuning is available at \url{https://github.com/ACEsuit/mace/tree/main}. The current implementation of multi-head fine-tuning, as well as the fine-tuning method that is being used in this work, allows the update of parameters of the common backbone. Performance of fine-tuning is demonstrated via \mlo~and \ml~. The same approach can be extended to other foundation models, which we plan to explore in future studies.

\section{RESULTS}
\label{sec:results}

In this section, we examine fine-tuning performance across various systems, including elemental metals (Section~\ref{sec:sub:metals}), silicon (Section~\ref{sec:sub:Si}), dislocations (Section~\ref{sec:sub:disloc}), high-entropy alloys (Section~\ref{sec:sub:HEA}), and the ionic crystal (Section~\ref{sec:sub:LiCl}). The evaluation covers mechanical properties, particularly in systems with point defects and dislocations, as well as selected thermodynamic and dynamic properties. For reference, we also compare fine-tuned models with those trained from scratch using the MACE and ACE architectures. Training details are provided in Section~\ref{sec:apd:training}.

\subsection{Elemental Metals}
\label{sec:sub:metals}

We conduct numerical experiments using the dataset from Zou et al.~\cite{Zuo2020}, a standard DFT dataset covering Li, Mo, Ni, Cu, Si, and Ge. Fine-tuning follows the ``FT-1" strategy (cf.~Section~\ref{sec:sub:ft}), selecting configurations directly from open-source datasets without filtering. These elements represent diverse chemistries (main group metals, transition metals, and semiconductors), crystal structures (BCC, FCC, and diamond), and bonding types (metallic and covalent). 

For evaluation, we use molecular statics and the climbing-image nudged elastic band (CI-NEB) method~\cite{henkelman2000climbing}. Key material properties—including cubic lattice parameter ($a_0$), elastic constants ($C_{11}$, $C_{12}$, $C_{44}$), bulk modulus ($B$), migration energy ($E_m$), vacancy formation energy ($E_v$), and activation barrier for vacancy diffusion ($E_a = E_v + E_m$)—are compared against DFT reference values. The analysis includes foundation models (\mlo~and \ml), their fine-tuned versions, and models trained from scratch (MACE and ACE).

This section focuses on Li, Cu, and Si, while results for the remaining elements are provided in the supplement (Section~\ref{sec:apd:numer}). Table~\ref{tab:metals} shows that the original foundation models exhibit large errors (20–80\%) for most properties, consistent with previous benchmarks~\cite{li2024extendable}, likely due to the low emphasis on stress during training. Fine-tuned models and those trained from scratch show significant improvements. Lattice parameters are predicted within 0.1–2.0\% of DFT values, and elastic constants typically within 10\%. While fine-tuned models (``Fine-tuning-0b" and ``Fine-tuning-0") achieve good accuracy for defect formation and migration energies in diamond systems, they tend to over- or underestimate vacancy formation energies and activation barriers in BCC and FCC systems, particularly compared to the ACE model trained from scratch.

\begin{table}[h!]
\centering
\resizebox{\textwidth}{!}{
\begin{tabular}{cccccccc}
  & DFT & \ml & Fine-tuning-0b & \mlo & Fine-tuning-0 & MACE-scratch & ACE-scratch \\
\hline
  & & & & \textbf{Li} & & & \\
  \hline
$a_0$ (\AA) & 3.427 & 3.399 (-0.8\%) & 3.451 (0.7\%) & 3.499 (2.1\%) & 3.479 (1.5\%) & 3.446 (0.5\%) & \textbf{3.441 (0.4\%)} \\
$C_{11}$ (GPa) & 15 & 19 (26.7\%) & 14 (-6.7\%) & 13 (-13.3\%) & 16 (6.7\%) & \textbf{15 (0.0\%)} & 18 (20.0\%) \\
$C_{12}$ (GPa) & 13 & \textbf{14 (7.7\%)} & \textbf{12 (-7.7\%)} & 11 (-15.4\%) & 15 (15.4\%) & 11 (-15.4\%) & 17 (30.8\%) \\
$C_{44}$ (GPa) & 11 & 13 (18.2\%) & \textbf{11 (0.0\%)} & 12 (9.1\%) & \textbf{11 (0.0\%)} & 10 (-9.1\%) & 9 (-18.2\%) \\
$B$ (GPa) & 14 & 16 (14.3\%) & \textbf{13 (-7.1\%)} & 12 (-14.3\%)  & \textbf{15 (7.1\%)} & 12 (-14.3\%) & 17 (21.4\%) \\
$E_v$ (eV) & 0.62 & 0.50 (-19.4\%) & 0.55 (-11.3\%) & 0.67 (8.1\%) & 0.65 (4.8\%) & \textbf{0.61 (-1.6\%)} & \textbf{0.61 (-1.6\%)} \\
$E_m$ (eV) & 0.06 & 0.15 (150.0\%) & 0.12 (100.0\%) & 0.11 (83.3\%) & \textbf{0.10 (66.7\%)} & 0.13 (116.7\%) & 0.14 (133.3\%) \\
$E_a$ (eV) & 0.68 & 0.65 (-4.4\%) & \textbf{0.68 (0.0\%)} & 0.78 (14.71\%) & 0.75 (10.2\%) & 0.74 (8.8\%) & 0.75 (10.3\%) \\
\hline \\


 & & & &  \textbf{Cu} & & & \\
\hline
$a_0$  &  3.621 &         3.615 (-0.2\%) &  \textbf{3.616 (-0.1\%)} &  \textbf{3.626 (+0.1\%)} &         3.634 (+0.4\%) &  3.643 (+0.6\%) &          3.632 (+0.3\%) \\
$C_{11}$ &    173 &  \textbf{171 (-1.2\%)} &    \textbf{171 (-1.2\%)} &             168 (-2.9\%) &           181 (+4.6\%) &    185 (+6.9\%) &            180 (+4.0\%) \\
$C_{12}$ &    133 &          119 (-10.5\%) &    \textbf{123 (-7.5\%)} &             120 (-9.8\%) &  \textbf{123 (-7.5\%)} &    121 (-9.0\%) &            122 (-8.3\%) \\
$C_{44}$ &     88 &           58 (-34.1\%) &             60 (-31.8\%) &             71 (-19.3\%) &   \textbf{86 (-2.3\%)} &     80 (-9.1\%) &             84 (-4.5\%) \\
$B$   &    146 &           136 (-6.8\%) &             138 (-5.5\%) &             135 (-7.5\%) &  \textbf{142 (-2.7\%)} &    140 (-4.1\%) &            141 (-3.4\%) \\
$E_v$  &   1.15 &         0.98 (-14.8\%) &           1.31 (+13.9\%) &           1.02 (-11.3\%) &         1.28 (+11.3\%) &  2.05 (+78.3\%) &  \textbf{1.07 (-7.0\%)} \\
$E_m$  &   0.79 &         0.63 (-20.3\%) &           0.69 (-12.7\%) &           0.62 (-21.5\%) &          0.73 (-7.6\%) &  0.62 (-21.5\%) &   \textbf{0.80 (+1.3\%)} \\
$E_a$ &   1.94 &         1.61 (-17.0\%) &    \textbf{2.00 (+3.1\%)} &           1.64 (-15.5\%) &          2.01 (+3.6\%) &  2.67 (+37.6\%) &           1.87 (-3.6\%) \\
\hline \\


  &  &  &  & \textbf{Si} & & &  \\
\hline
$a_0$  &  5.469 &          5.493 (+0.4\%) &           5.456 (-0.2\%) &        5.455 (-0.3\%) &  5.456 (-0.2\%) &  \textbf{5.467 (-0.0\%)} &          5.463 (-0.1\%) \\
$C_{11}$ &    156 &            92 (-41.0\%) &            123 (-21.2\%) &         102 (-34.6\%) &   123 (-21.2\%) &             93 (-40.4\%) &  \textbf{144 (-7.7\%)} \\
$C_{12}$ &     65 &            45 (-30.8\%) &             55 (-15.4\%) &  \textbf{61 (-6.2\%)} &     59 (-9.2\%) &             50 (-23.1\%) &            90 (+38.5\%) \\
$C_{44}$ &     76 &            34 (-55.3\%) &             59 (-22.4\%) &          31 (-59.2\%) &    62 (-18.4\%) &             45 (-40.8\%) &  \textbf{73 (-3.9\%)} \\
$B$   &     95 &            61 (-35.8\%) &             73 (-23.2\%) &          74 (-22.1\%) &    75 (-21.1\%) &             68 (-28.4\%) &  \textbf{108 (+13.7\%)} \\
$E_v$  &   3.25 &          1.26 (-61.2\%) &  \textbf{2.98 (-8.3\%)} &        1.76 (-45.8\%) &  2.41 (-25.8\%) &           2.29 (-29.5\%) &          2.87 (-11.7\%) \\
$E_m$  &   0.21 &  \textbf{0.23 (+9.5\%)} &           0.07 (-66.7\%) &        0.12 (-42.9\%) &  0.25 (+19.0\%) &            0.10 (-52.4\%) &          0.02 (-90.5\%) \\
$E_a$ &   3.46 &          1.49 (-56.9\%) &  \textbf{3.05 (-11.8\%)} &        1.88 (-45.7\%) &  1.88 (-45.7\%) &           2.66 (-23.1\%) &          2.89 (-16.5\%) \\
\hline \\

\end{tabular}
}
\caption{Comparison of values and percentage errors for selected materials properties (cubic lattice parameter $a_0$, elastic constants ($C_{11}, C_{12}$ and $C_{44}$), bulk modulus $B$, migration energy ($E_m$), and vacancy formation energy ($E_v$) as well as activation barrier for vacancy diffusion ($E_a = E_v + E_m$) obtained using a range of MLIPs (pre-trained models \ml~and~\mlo, their corresponding fine-tuned models, and two training from scratch models (MACE and ACE) with respect to DFT reference values from~\cite{Zuo2020}.}
\label{tab:metals} 
\end{table}

Our results indicate that the limitations of foundation models in predicting the elastic properties of pure metals likely arise from insufficient stress data or a lack of emphasis on stress during training. Fine-tuning significantly improves these predictions, bringing them closer to DFT-level accuracy. However, on smaller or simpler datasets, it may offer little advantage over training from scratch. We anticipate that using richer datasets for fine-tuning will further mitigate these limitations, as demonstrated in the following numerical examples.

\subsection{A Typical Semiconductor: Silicon}
\label{sec:sub:Si}

We use silicon with a more complex dataset to evaluate the impact of dataset selection. The original database, curated based on domain expertise, provides broad coverage of material properties~\cite{bartok2018machine}. It contains 531,710 electronic-structure data points, including total energy, stress tensor components, and force components.

To examine the role of dataset selection in fine-tuning, we compare two strategies. The first, ``FT-1," utilizes the full dataset, while the second, ``FT-2," selects a subset focusing on defect configurations and low-temperature states, particularly relevant for mechanical properties~\cite{toit2024hyperparameter}.  

\begin{figure}
\centering
\includegraphics[height=6.5cm]{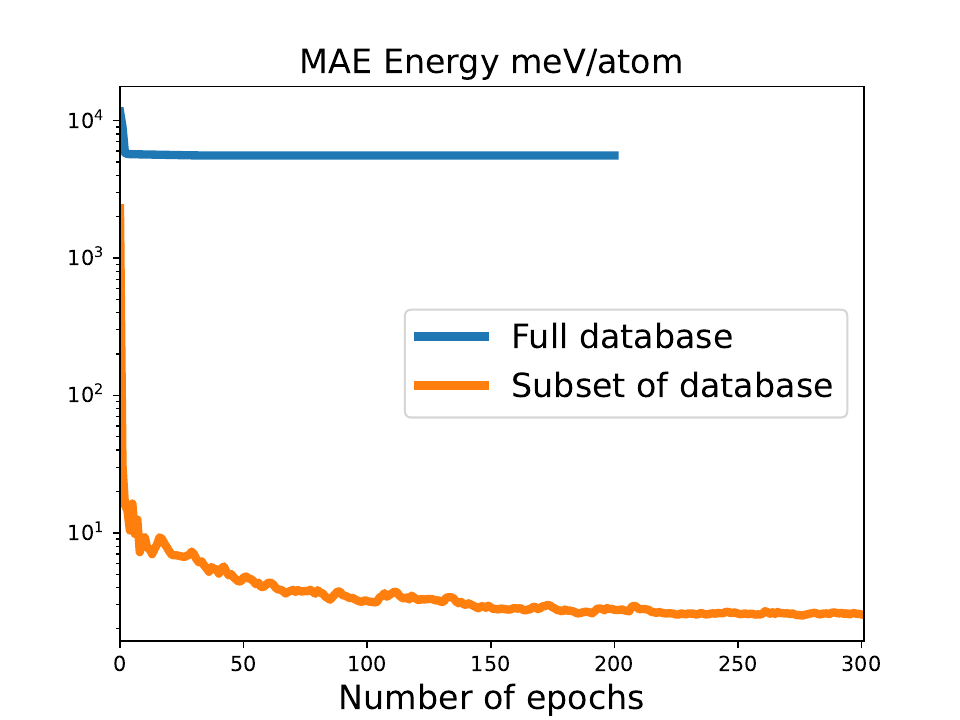}
\includegraphics[height=6.5cm]{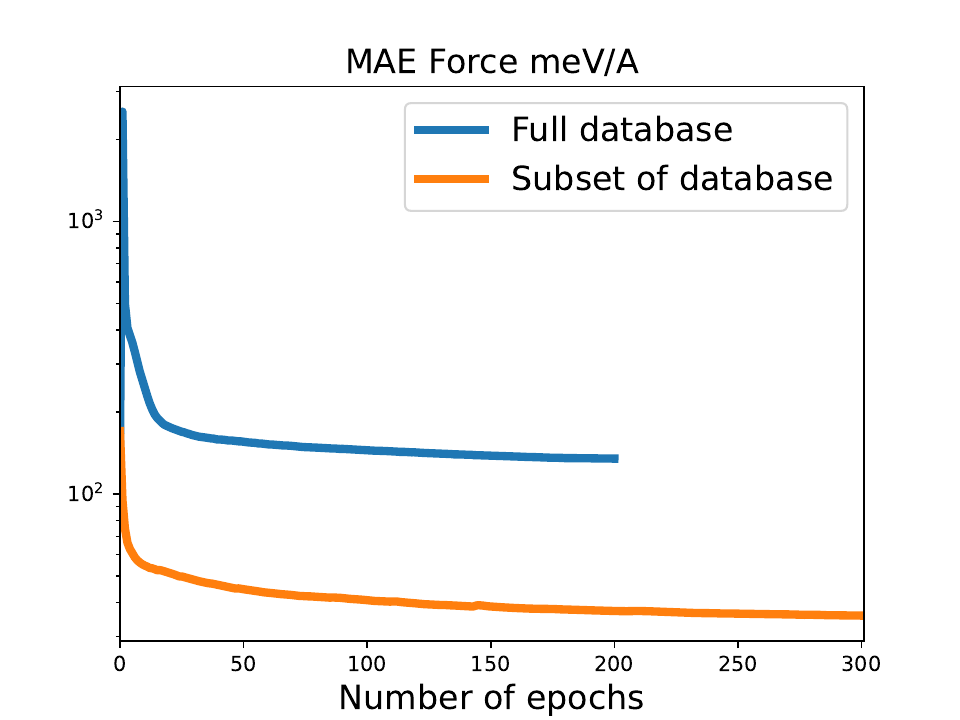}
\caption{Testing MAE versus number of epochs during fine-tuning of the foundation model~\ml.}
\label{fig:Si-RMSE}
\end{figure}

Figure~\ref{fig:Si-RMSE} presents the mean absolute error (MAE) for energy and forces during fine-tuning for both dataset strategies. Fine-tuning with the full dataset struggles to converge, with energy and force errors plateauing at high values, especially in energy predictions. In contrast, fine-tuning with the selected subset leads to better convergence, with final RMSE values of 7 meV per atom for energy and 60 meV/\AA~for forces. This suggests that parts of the full dataset may exhibit significant deviations from the reference DFT calculations, possibly due to the presence of amorphous or poorly defined configurations. When such configurations are excluded, and a cleaner subset is used for fine-tuning, the training process becomes more stable and yields improved results. These results underscore the importance of careful dataset selection when fine-tuning foundation models. Strategies such as filtering~\cite{hyperactive2022} or active learning~\cite{zhang2023atomistic} can help construct {\it quasi-optimal} fine-tuning datasets while reducing the computational cost of DFT calculations. Based on these findings, subsequent tests in this section will focus on results obtained using the ``FT-2" strategy.

\subsubsection{Mechanical properties}
\label{sec:sub:sub:basic}

We first evaluate the mechanical properties, following a similar procedure as in previous sections. Table~\ref{tab:Si-elastic} summarizes results for various MLIPs, including foundation models (\ml~and~\mlo), their fine-tuned versions, and models trained from scratch (MACE and ACE).  

\begin{figure}
\centering
\includegraphics[height=8cm]{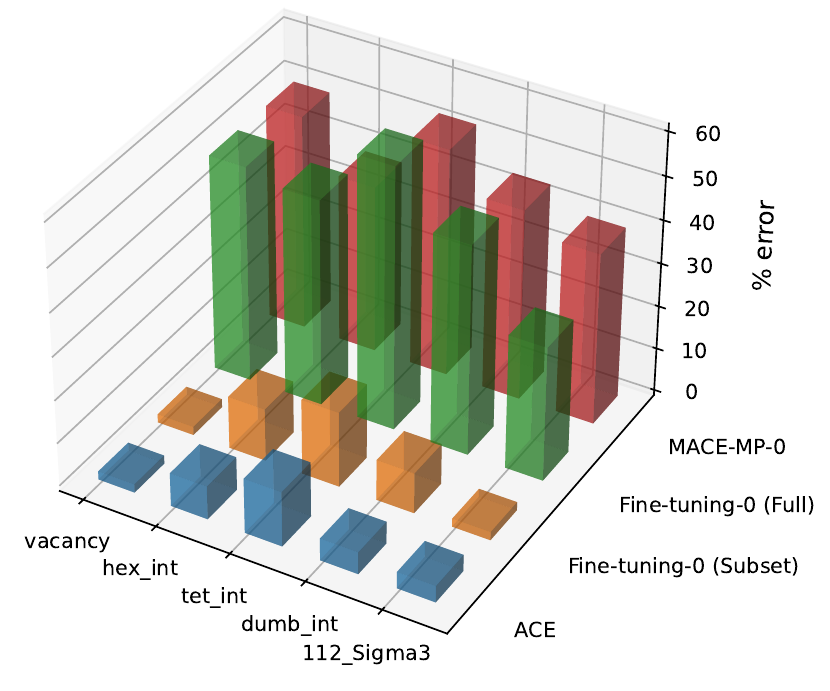}
\caption{Relative error in defect formation energy of silicon for various MLIPs, including the pre-trained model (MACE-MP-0) with fine-tuning on the full and subset training sets, without fine-tuning, and the ACE model trained from scratch.}
\label{fig:Si-prop}
\end{figure}

For bulk properties, the fine-tuned \ml~model achieves the best performance, with two properties within 1\% of DFT values. In surface energy predictions, the fine-tuned \mlo~model excels for the (111) and (110) cleavage planes, while the ACE model, trained from scratch, performs best overall, with errors around 3\% relative to DFT. Formation energies of vacancy and interstitial defects show varying errors, though most predictions remain within 10\% of DFT references. Fine-tuning generally improves accuracy across different properties. Additionally, we assess the (112) $\Sigma_3$ symmetric tilt grain boundary, a common planar defect in silicon, with results consistent with those for point defects. Figure~\ref{fig:Si-prop} visualizes these findings, highlighting the improvements achieved through fine-tuning.  

\begin{table}[h!]
\centering
\resizebox{\textwidth}{!}{
\begin{tabular}{cccccccc}
 & DFT & \ml & Fine-tuning-0b & \mlo & Fine-tuning-0 & MACE-scratch & ACE-scratch \\
\hline
$C_{11}$ & 153.3 & 92.2 (-39.8\%) & 145.3 (-5.2\%) & 102.5 (-33.2\%) & 145.5 (-5.1\%) & 132.6 (-13.5\%) & \textbf{151.1 (-1.5\%)} \\
$C_{12}$ & 56.3 & 45.6 (-19.1\%) & \textbf{55.7 (-1.1\%)} & 60.6 (7.6\%) & 57.0 (1.3\%) & 53.9 (-4.2\%) & 56.9 (1.2\%) \\
$C_{44}$ & 72.2 & 34.1 (-52.8\%) & \textbf{71.8 (-0.6\%)} & 30.9 (-57.1\%) & 70.7 (-2.1\%) & 73.6 (2.0\%) & 71.3 (-1.2\%) \\
$B$ & 88.6 & 61.1 (-31.0\%) & 84.0 (-5.2\%) & 74.5 (-15.9\%) & 86.6 (-2.3\%) & 86.3 (-2.6\%) & \textbf{88.3 (-0.3\%)} \\
$(111)$ & 1.57 & 0.64 (-59.0\%) & 1.49 (-5.1\%) & 0.75 (-52.2\%) & 1.49 (-5.3\%) & 1.44 (-8.3\%) & \textbf{1.51 (-3.8\%)} \\
$(110)$ & 1.52 & 0.85 (-43.8\%) & 1.47 (-3.2\%) & 0.92 (-39.4\%) & 1.48 (-2.6\%) & 1.45 (-4.9\%) & \textbf{1.51 (-0.5\%)} \\
vac. & 3.67 & 1.46 (-60.3\%) & 3.54 (-3.5\%) & 1.85 (-49.6\%) & 3.51 (-4.3\%) & 3.51 (-4.3\%) & \textbf{3.63 (-0.9\%)} \\
hex. int. & 3.72 & 1.79 (-51.9\%) & 3.41 (-8.3\%) & 2.28 (-38.6\%) & \textbf{3.87 (4.1\%)} & 3.47 (-6.8\%) & 3.54 (-4.7\%) \\
tet. int. & 3.91 & 1.70 (-56.5\%) & 3.62 (-7.3\%) & 1.86 (-52.5\%) & \textbf{3.64 (-6.9\%)} & 3.51 (-10.1\%) & 3.61 (-7.7\%) \\
dumb. int. & 3.66 & 1.82 (-50.2\%) & 3.42 (-6.2\%) & 2.04 (-44.1\%) & \textbf{3.58 (-2.0\%)} & 3.39 (-7.2\%) & 3.58 (-2.1\%) \\
(112) $\Sigma_3$ & 0.93 & 0.45 (-51.2\%) & 0.87 (-6.5\%) & 0.57 (-38.5\%) & \textbf{0.96 (3.5\%)} & 0.88 (-4.8\%) & 0.89 (-3.8\%) \\
\end{tabular}
}
\caption{Comparison of values and percentage errors for defects formation energies obtained using a range of interatomic potentials, with respect to DFT reference values from the literature~\cite{bartok2018machine}.}
\label{tab:Si-elastic} 
\end{table}

\subsubsection{Di-interstitial}
\label{sec:sub:sub:diinterstitial}

Next, we analyze the di-interstitial case, which was also investigated in the original work by Bartók et al.~\cite{bartok2018machine}. This test evaluates the transferability of the models to new defect types, as the atomic environments in di-interstitial configurations differ significantly from those in the training dataset. Table~\ref{tab:Si-di} presents the percentage error in the formation energy of the di-interstitial across six conformations, as predicted by various interatomic potentials. The results show that the original foundation models struggle to accurately predict this property. In contrast, models trained from scratch, such as MACE and ACE, perform better, achieving accuracy within approximately 10\%. Fine-tuned foundation models (``Fine-tuning-0b" and ``Fine-tuning-0") further improve these predictions, with the \mlo~model achieving relative errors within 10\%. These findings align with trends observed in the formation energy of single point defects, as shown in Table~\ref{tab:Si-elastic}.

\begin{table}[h!]
\centering
\resizebox{\textwidth}{!}{
\begin{tabular}{ccccccc}
 & \ml & Fine-tuning-0b & \mlo & Fine-tuning-0 & MACE-scratch & ACE-scratch \\
\hline
EXT & -51.0\% & -6.3\% & -40.5\% & -4.3\% & -2.5\% & \textbf{1.2\%} \\
TT & -62.4\% & -12.3\% & -54.7\% & \textbf{-9.8\%} & -10.1\% & -15.6\% \\
XEX & -52.5\% & -6.5\%  & -43.3\% & -6.7\% & -9.3\% & \textbf{-5.5\%} \\
XT & -52.8\% & -10.0\%  & -49.1\% & \textbf{-6.5\%} & -7.6\% & -8.9\% \\
W & -48.0\% & -7.2\% & -40.8\% & -2.0\% & -4.6\% & \textbf{1.2\%} \\
XX3 & -53.8\% & -7.1\% & -44.8\% & \textbf{-4.5\%} & -5.2\% & -4.6\% \\
\end{tabular}
}
\caption{Comparison of values and percentage errors for the formation energy of di-interstitials in various configurations obtained using a range of interatomic potentials, with respect to DFT reference values from the literature~\cite{bartok2018machine}.}
\label{tab:Si-di} 
\end{table}

\subsection{Dislocations for BCC Metals}
\label{sec:sub:disloc}

Dislocations are critical to the mechanical properties of BCC metals, influencing strength and ductility. Understanding their behavior is key to predicting material performance under stress and guiding material design. Without fine-tuning, the foundation model \mlo~performs poorly on BCC metals~\cite{batatia2023foundation}, with defect formation energies and dislocation glide properties showing relative errors exceeding 50\% compared to DFT calculations.

To improve accuracy, we fine-tuned \mlo~and~\ml~foundation models for BCC metals and tested it on iron and tungsten, two well-studied systems in MLIPs literature. Public datasets (``FT-1” strategy) were used for benchmarking. In BCC materials, dislocations primarily occur in the $\langle111\rangle\{110\}$ and $\langle100\rangle\{010\}$ slip systems. We analyzed $\langle111\rangle$ screw and $\langle100\rangle$ edge dislocations by calculating transition pathways and Peierls barriers using the NEB method~\cite{henkelman2000climbing}.

\subsubsection{Iron}
\label{sec:sub:sub:iron}

The first BCC example we examine is iron (Fe), using a dataset constructed through active learning as proposed in~\cite{zhang2023atomistic}. This dataset is designed to capture deformation mechanisms at the crack tip, effectively illustrating the semi-brittle nature of BCC iron, making it well-suited for fine-tuning tasks involving extended defects.

For dislocation system tests, the simulation cells contained approximately 1,400 atoms for screw and 2,200 atoms for edge dislocations. Initial configurations were obtained through geometry optimization using the FIRE algorithm~\cite{guenole2020assessment}, with a force tolerance of $1\times10^{-6}$ eV/Å. The \texttt{matscipy.dislocation} module~\cite{grigorev2024matscipy} was used to generate and analyze atomistic dislocation structures. Transition path calculations were performed with an adaptive ODE solver, using eleven intermediate images and a stopping force tolerance of 0.025 eV/Å. The initial NEB relaxation paths were generated via linear interpolation between initial and final configurations.

\begin{figure}
\centering
\includegraphics[height=6.54cm]{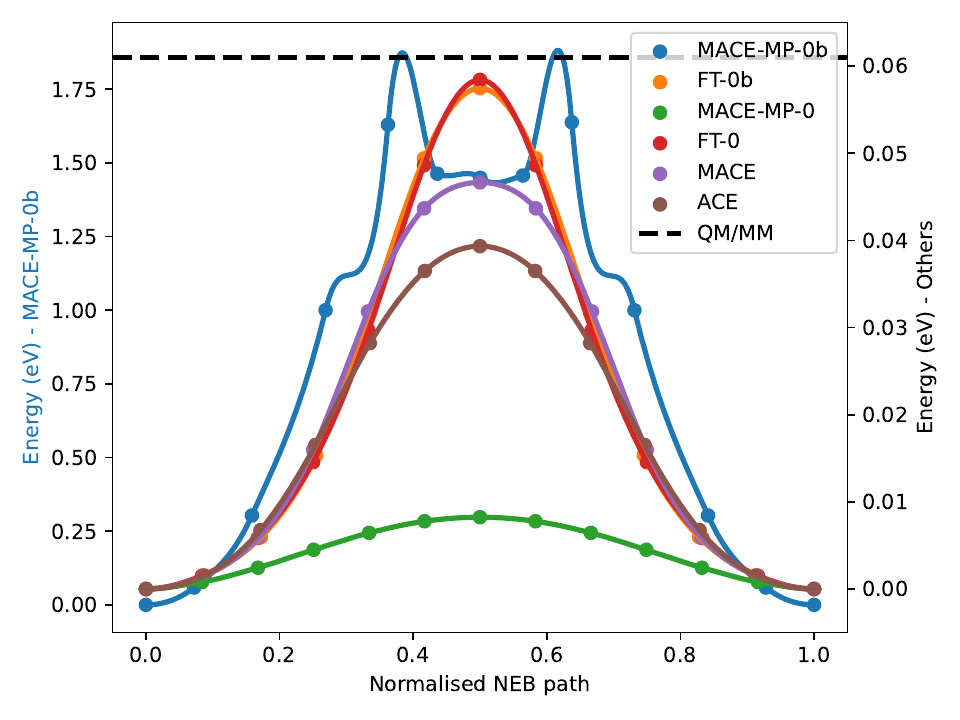}
\includegraphics[height=6.5cm]{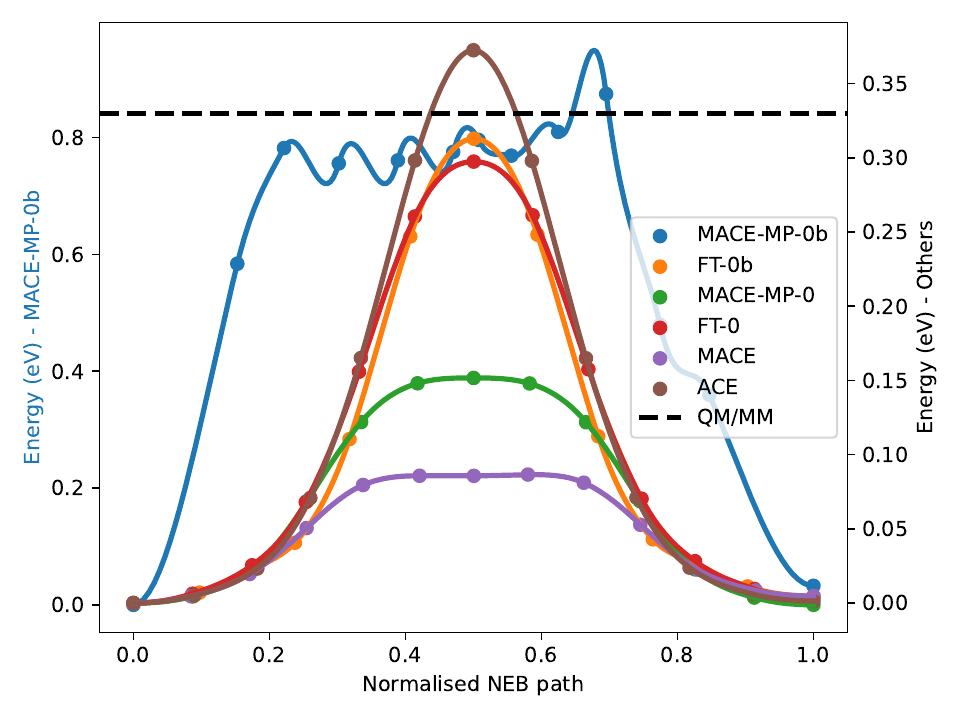}
\caption{Dislocation glide energy barriers in Fe calculated using various MLIPs, including two foundation models (with and without fine-tuning) and two models trained from scratch (ACE and MACE). Results are compared to the reference QM/MM calculation (black dashed line for the energy barrier). Left: $\langle111\rangle$ screw dislocation; Right: $\langle100\rangle$ edge dislocation.}
\label{fig:Fe}
\end{figure}

Figure~\ref{fig:Fe} shows the NEB minimum energy path, comparing Peierls barriers for $\langle111\rangle$ screw and $\langle100\rangle$ edge dislocations in Fe against QM/MM results. For screw dislocation (the left panel in Figure~\ref{fig:Fe}), the original foundation model (\mlo) captures the correct barrier shape but significantly underestimates its height, consistent with previous findings~\cite{batatia2023foundation}. The variant model (\ml) predicts an incorrect shape with a misplaced saddle point. Models trained from scratch (MACE and ACE) perform better, as their training datasets include defect structures, though they slightly underestimate the barrier. Fine-tuned foundation models (``FT-0b" and ``FT-0") yield the best results, closely matching the QM/MM predictions in both shape and barrier height. For edge dislocation (the right panel in Figure~\ref{fig:Fe}), the trends are similar. The MACE model trained from scratch predicts the lowest barrier, while the ACE model predicts the highest. Fine-tuned models again provide the most accurate results, aligning well with reference calculations.

Overall, fine-tuning significantly improves agreement with QM/MM results. The original foundation model struggles with dislocation slip due to the absence of defect configurations or insufficient stress training. However, it still serves as a strong starting point, enabling accurate predictions once fine-tuned with appropriate defect data.

\subsubsection{Tungsten}
\label{sec:sub:sub:tungsten}

The numerical setup for tungsten dislocations follows a similar approach. The dataset, obtained from a public repository~\cite{byggmastar2019machine}, already includes essential dislocation configurations. Therefore, the ``FT-1” strategy was used for fine-tuning.

\begin{figure}
\centering
\includegraphics[height=6.5cm]{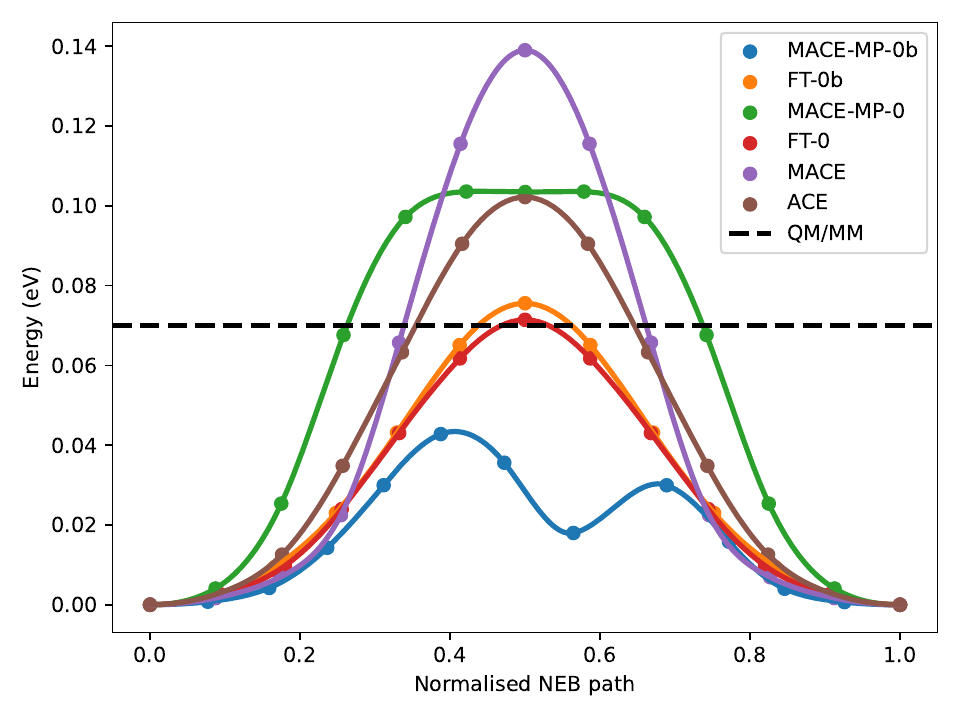}
\includegraphics[height=6.5cm]{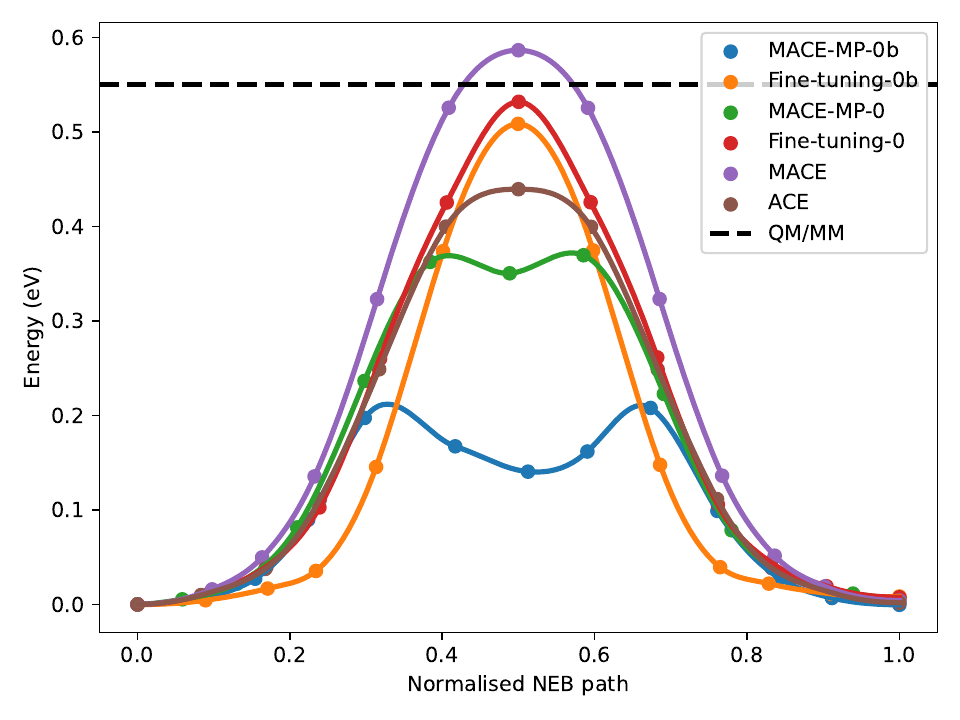}
\caption{Dislocation glide energy barriers in W calculated using various MLIPs, including two pre-trained models (with and without fine-tuning) and two models trained from scratch (ACE and MACE). Results are compared to the reference QM/MM calculation (black dashed line). Left: $\langle111\rangle$ screw dislocation; Right: $\langle100\rangle$ edge dislocation.}
\label{fig:W}
\end{figure}

Figure~\ref{fig:W} shows the NEB minimum energy path, comparing Peierls barriers for $\langle111\rangle$ screw and $\langle100\rangle$ edge dislocations in W. The results are consistent with those for iron in Figure~\ref{fig:Fe}. Both foundation models struggle to accurately capture the saddle points, leading to deviations in the NEB path. Training from scratch improves the predictions but remains suboptimal. In contrast, fine-tuned foundation models (``FT-0b" and ``FT-0") yield the best results, closely matching DFT energy barriers.

Across both iron and tungsten, fine-tuned foundation models outperform those trained from scratch, achieving faster convergence. This is likely because fine-tuning leverages a strong initial guess from the foundation model, requiring only minor adjustments instead of learning from scratch.

\subsection{A Multi-Components System: High-Entropy Alloy}
\label{sec:sub:HEA}

High-Entropy alloys (HEAs) have attracted significant attention due to their unique mechanical properties and exceptional resilience in harsh environments~\cite{miracle2017critical, george2019high}. Here, we demonstrate that fine-tuning foundation models improves performance not only in single-element systems but also in complex multi-element alloys.

This study focuses on the NbAlMoTiZr HEA system. The dataset was generated using the ``FT-3” approach, which first constructs a large candidate dataset and then applies the filtering strategy from~\cite{hyperactive2022} to obtain the final set. The candidate dataset includes configurations with 1–5 atoms, for which DFT energies and forces were computed, while configurations with 6 atoms were randomly selected for testing.

We observe that fine-tuning not only improves predictive accuracy but also accelerates learning, as seen in the relationship between RMSE and training epochs. This advantage stems from the strong initial guess provided by foundation models, reducing the training effort needed to reach high accuracy. As shown in Table~\ref{fig:error-hea}, fine-tuned foundation models consistently outperform models trained from scratch, including ACE and MACE, highlighting the efficiency of fine-tuning. This allows users to achieve high accuracy with minimal computational cost and without extensive expertise in the MLIPs framework.

\begin{table}[h!]
\centering
\resizebox{0.6\textwidth}{!}
{
\begin{tabular}{ccc}
 & MAE Energy (meV/atom) & MAE Force (meV/\AA) \\
\hline
\ml & 63.8 & 132.5 \\
\mlo & 58.6 & 102.3 \\
Fine-tuning-0b & \textbf{13.8} & \textbf{20.3} \\
Fine-tuning-0 & \textbf{13.8} & \textbf{20.1} \\
MACE & 16.4 & 23.2 \\
ACE & 24.1 & 86.2 \\
\end{tabular}
}
\caption{Testing RMSE error for NbAlMoTiZr high-entropy alloys. Fine-tuned models outperform all other models, with the best results highlighted.}
\label{fig:error-hea}
\end{table}

\begin{figure}
\centering
\includegraphics[height=6.5cm]{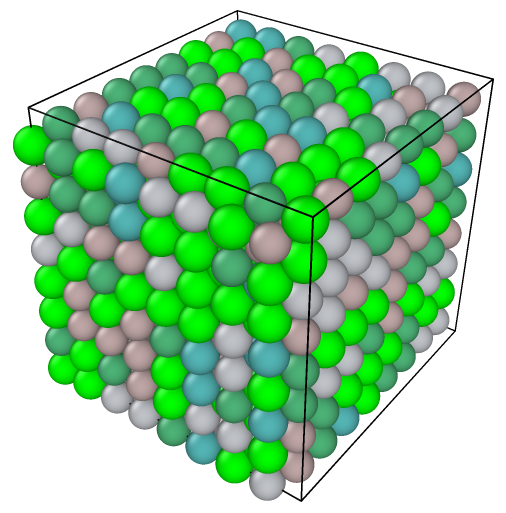}\qquad
\includegraphics[height=6.5cm]{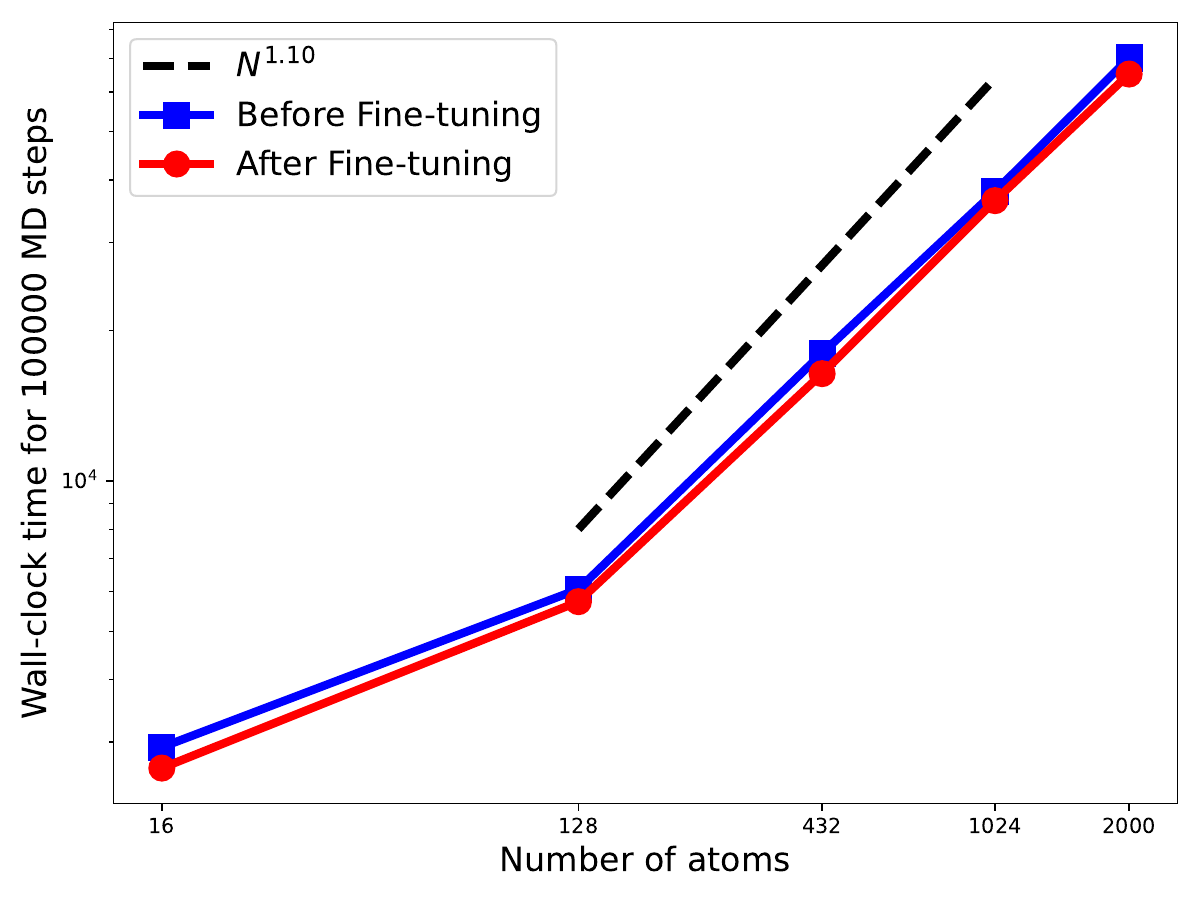}
\caption{Left: Illustration of the NbAlMoTiZr high-entropy alloy, colored by element using Ovito software \cite{stukowski2009visualization}; Right: Scaling and computational cost analysis of MD simulations before and after fine-tuning for NbAlMoTiZr high-entropy alloys.}
\label{fig:hea-illus}
\end{figure}

To evaluate the scalability of our fine-tuning approach and explore potential improvements, we perform MD simulations on supercells of varying sizes using the \ml~model before and after fine-tuning. The left panel of Figure~\ref{fig:hea-illus} presents an example configuration with 512 atoms from the high-entropy alloy system.

The right panel of Figure~\ref{fig:hea-illus} presents the scaling and computational cost analysis for MD simulations of NbAlMoTiZr high-entropy alloys, showing running time for 100,000 steps. These simulations benchmark runtime and computational efficiency across different system sizes. To ensure reproducibility, average runtimes are reported based on five trials per configuration.

The results indicate that fine-tuning significantly improves model accuracy but has minimal impact on computational efficiency or scaling. This suggests that while fine-tuning enhances predictive performance, it does not substantially reduce model size, highlighting the potential need for model distillation~\cite{gou2021knowledge, alkhulaifi2021knowledge} to optimize the foundation model further. Additionally, while this study focuses on RMSE accuracy, future work will explore other key properties of high-entropy alloys, such as diffusion and thermal conductivity, using various atomistic foundation models.




\subsection{An Ionic Crystal: LiCl}
\label{sec:sub:LiCl}

LiCl is extensively studied for its unique ionic properties, which influence key characteristics such as conductivity and thermal stability. Understanding its atomic structure and behavior under various conditions is essential for applications in energy storage and electrochemical systems, where precise control of material properties is crucial.

In this work, we fine-tune a foundation model using the dataset from~\cite{sivaraman2021automated} (``FT-1" strategy) within the MD simulation setup described in the same study. The original work employs dispersion-corrected DFT single-point calculations using the PBE exchange-correlation functional with ``D3" dispersion correction and the projector-augmented wave method. Therefore, no additional dispersion correction is required for the fine-tuned model.

We focus on evaluating the fine-tuned variant of the \ml~model, as the original unrefined model already yields comparable results. Fine-tuning enables a systematic assessment of its enhanced predictive capabilities for structural and transport properties of molten LiCl, demonstrating how fine-tuning improves the accuracy of large pre-trained models in capturing the complex behavior of molten salts. To maintain conciseness, the radial distribution function (RDF) of molten LiCl at 900 K is presented later in Section~\ref{sec:apd:numer}~(Figure~\ref{fig:LiCl-rdf}).

\begin{figure}
\centering
\includegraphics[height=6.5cm]{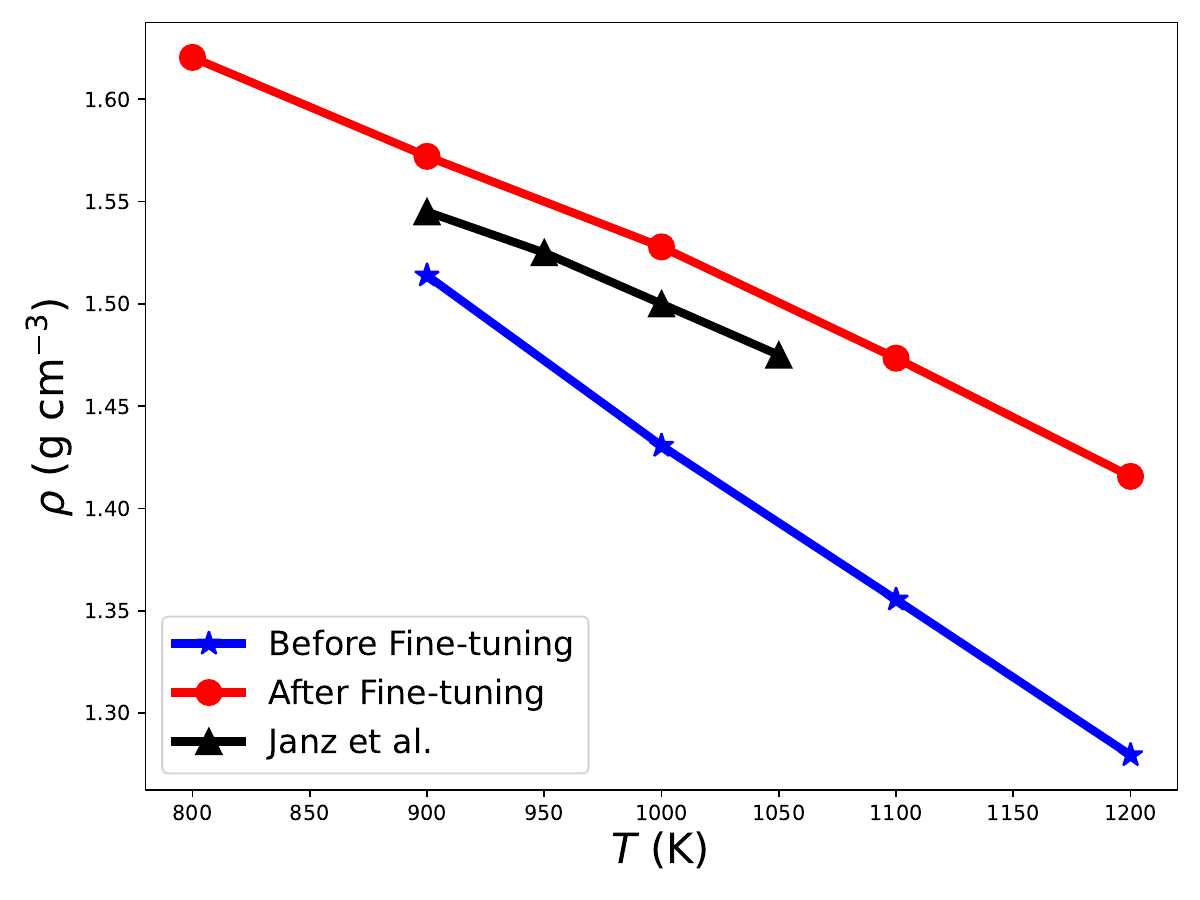}
\includegraphics[height=6.5cm]{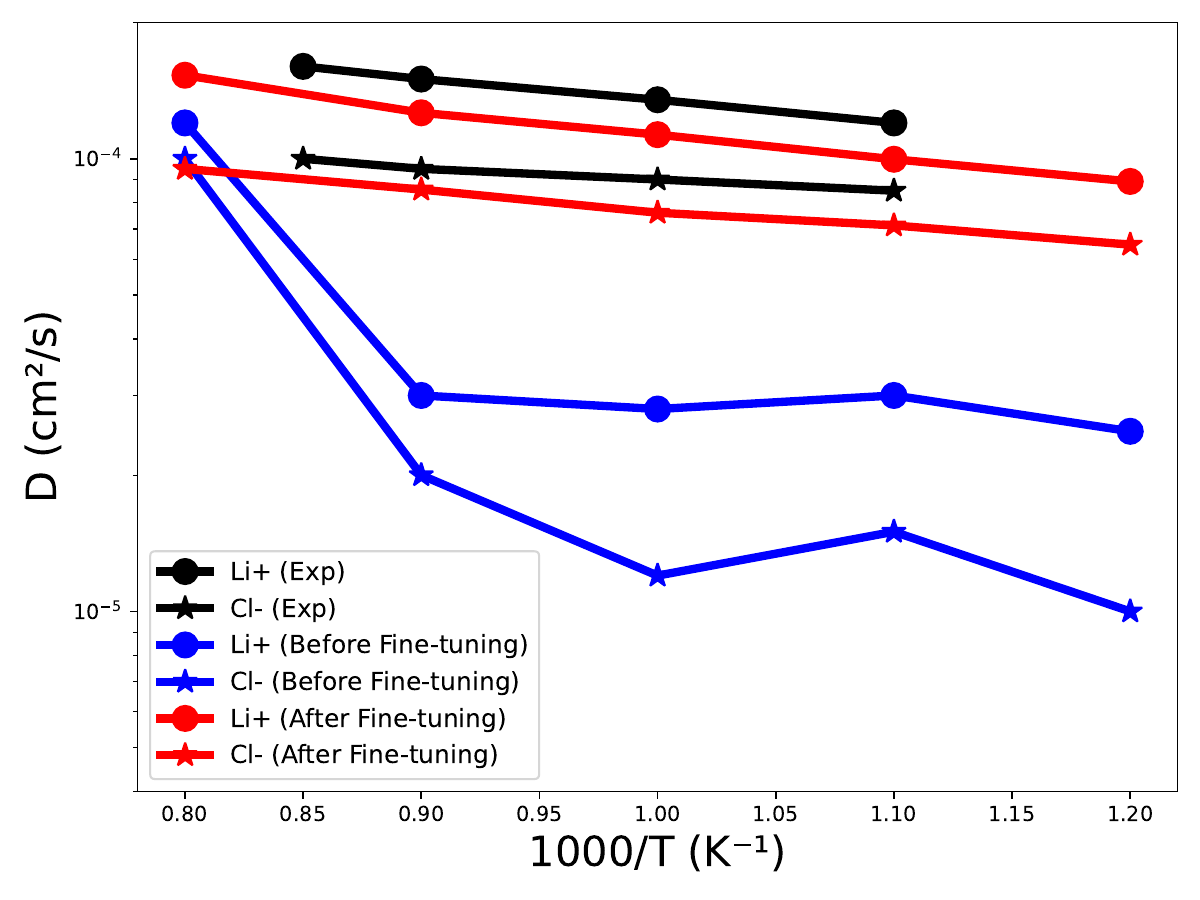}
\caption{Left: Density of molten LiCl at various temperatures, comparing results from \ml~and its fine-tuned MD simulation. Right: Self-diffusion coefficients for $\text{Li}^+$, $\text{Cl}^-$, comparing results from \ml~and its fine-tuned MD simulation. Dotted lines are the linear fit between the log scale diffusion coefficient and the reciprocal of temperature.}
\label{fig:LiCl-density}
\end{figure}

The left panel of Figure~\ref{fig:LiCl-density} compares the density ($\rho$) of molten LiCl over a range of temperatures for the \ml~model before (blue solid line) and after fine-tuning (red solid line), with experimental data from~\cite{sivaraman2021automated} (black solid line) as a reference. Before fine-tuning, the simulated density consistently underestimates experimental values across all temperatures. After fine-tuning, the model shows significantly improved agreement, reducing errors and accurately capturing the characteristic linear decrease in density with increasing temperature.

The right panel of Figure~\ref{fig:LiCl-density} shows the self-diffusion coefficients ($D$) of $\text{Li}^+$ and $\text{Cl}^-$ ions in molten LiCl, plotted against $1000/T$ (the reciprocal temperature). Results are compared for three cases: experimental values (black dotted lines with symbols), \ml~predictions before fine-tuning (blue solid lines with symbols), and after fine-tuning (red solid lines with symbols). Before fine-tuning, the diffusion coefficients deviate significantly from experimental values, with $\text{Li}^+$ being overestimated and $\text{Cl}^-$ underestimated across all temperatures. Fine-tuning substantially improves agreement with experimental trends, particularly refining the relative magnitudes and slopes of the diffusion coefficients.



This case highlights the effectiveness of fine-tuning atomistic foundation models for molten LiCl. The fine-tuned potential improves structural and transport property predictions. Future work will extend this approach to complex ionic systems and refine optimization strategies for greater accuracy.

\section{SUMMARY AND DISCUSSIONS}
\label{sec:conclusion}

In this work, we evaluate the fine-tuning performance of the recently developed atomistic foundation models, \mlo~and its variant \ml, with a focus on their accuracy under varying computational settings. Notably, the fine-tuning process of U-MLIPs can be efficiently performed using the RBMD package. To facilitate reproducibility and broader adoption, we provide an open-access platform at \url{https://www.randbatch.com/rbmd}, which hosts example workflows and implementation scripts for realistic simulation scenarios, with continued updates planned. Beyond benchmarking foundation and fine-tuned models, we also investigate training from scratch, yielding two key insights:

\begin{itemize}
\item \textit{Fine-tuning vs. Training from Scratch:} Fine-tuning leverages the knowledge embedded in foundation models, enabling faster convergence and reducing data requirements compared to training from scratch. While training from scratch offers full control, it demands significantly larger datasets. In contrast, fine-tuning efficiently adapts a general-purpose model to specific applications, making it particularly valuable for complex systems where data generation is costly. Our results show that fine-tuning improves accuracy across diverse material systems, including metals, ionic compounds, and high-entropy alloys, while reducing training effort.

\item \textit{Dataset Selection for Fine-tuning:} Training data selection is crucial for successful fine-tuning. Using defect-rich configurations and physically relevant structures improves accuracy and generalizability, as observed in our studies on metallic systems and ionic compounds. Aligning datasets with the target distribution enhances transferability, while uncertainty quantification helps filter high-variance samples, refining the training process and ensuring robust performance in complex material systems.

\end{itemize}

For the continued development and application of atomistic foundation models, we propose the following key directions:

\begin{itemize}

\item \textit{Improving Fine-tuning Methods:} Fine-tuning techniques can be further refined by incorporating more advanced machine learning approaches, such as DoRA~\cite{liu2024dora}, which improves representation disentanglement and adaptation~\cite{kim2024hydra}. In addition, a better understanding of the mathematical foundations of fine-tuning is needed to establish a more solid theoretical basis. A combination of rigorous theory and practical fine-tuning strategies will be important for improving the reliability and efficiency of foundation models.

\item \textit{Meta-learning and Transfer Learning:} Meta-learning and transfer learning could make foundation models more feasible for complicated applications. Meta-learning helps models learn efficiently from limited data, while transfer learning allows knowledge from one dataset to be applied to others, improving generalization. These techniques have been effective in other fields and could also be valuable for MLIPs. Recent studies~\cite{allen2024learning} have explored this direction.

\item \textit{Extending to Complex Systems and Long-time Simulations:} Applying MLIPs to complex systems and long-time dynamics remains a challenge. Testing foundation models on such systems is computationally expensive, and ensuring accuracy and stability over long simulations is difficult. These issues are particularly relevant for materials science, where long-timescale behavior is key to understanding many properties. Advances in high-performance computing will be necessary to make large-scale applications more feasible. Some recent work~\cite{matin2025teacher} has begun addressing these challenges.

\item \textit{Integrating Foundation Models into Multiscale Methods:} Foundation models and their fine-tuned versions have the potential to enhance the accuracy and efficiency of multiscale simulations by bridging different length and time scales. It is interesting to explore the direct integration of foundation models into multiscale frameworks, such as QM/MM methods~\cite{chen2022qm, wang2021posteriori}, adaptive mesh refinement strategies~\cite{fu2025meshac, fu2023adaptive}, and concurrent atomistic-continuum approaches~\cite{fu2025meshac, wang2025posteriori}. Integrating foundation models into multiscale methods allows for better uncertainty estimation and error control, improving the reliability of simulations for complex material systems.

\item \textit{Developing Foundation Models for Specific Fields:} Another direction is developing foundation models designed for specific classes of materials, such as transition metal dichalcogenide alloys~\cite{siddiqui2024machine}, metal-organic frameworks~\cite{vandenhaute2023machine}, battery materials~\cite{harper2020ab}, and low-dimensional materials~\cite{wang2022anisotropic}. Tailoring models to particular material types can improve accuracy and applicability, complementing more general-purpose atomistic foundation models. 

\end{itemize}





\section{SUPPLEMENTARY INFORMATION}

\subsection{Training Details}
\label{sec:apd:training}

In this section, we outline the procedures for both fine-tuning and training models from scratch. The MACE architecture is implemented in PyTorch~\cite{paszke2019pytorch} and employs the \texttt{e3nn} library~\cite{geiger2022e3nn}. The MACE training and evaluation codes are distributed via GitHub under the MIT license, available at \url{https://github.com/ACEsuit/mace/}.  Two atomistic foundation models are available at \url{https://github.com/ACEsuit/mace-mp/}, with the MACE training and evaluation codes distributed under the MIT license at \url{https://github.com/ACEsuit/mace/}. 

The fine-tuning procedure is performed using the Random-Batch Molecular Dynamics (RBMD) package, which enables fast simulations of particle systems at the nano/micro scale. Unlike existing packages, RBMD employs random batch methods~\cite{RBM_Jin, RBE, RBE_2, IRBE} for handling nonbonded interactions in particle systems. This approach allows simulations of up to 10 million particles on a single GPU with a CPU core~\cite{gao2024rbmd}. The platform for fine-tuning the MACE foundation model is accessible at \url{https://www.randbatch.com/rbmd}. All data and scripts supporting this work will also be made publicly available at the same platform.

For fine-tuning the two foundation models (\mlo~and~\ml), we trained them using energy, force, and stress labels (if available) with a 1-10-100 loss ratio under the mean squared error (MSE) criterion. Since we used the medium models, we set the maximal message equivariance to $L=1$. Key model parameters, such as correlation order in each layer, radial cutoff, number of hidden layers and units, and activation function, were kept at their default MACE settings. The training data, derived from well-established DFT ab-initio sources, were split with a 20:1 training-validation ratio. We employed the Adam optimizer~\cite{kingma2014adam} with a learning rate of 5e-4 over 200 epochs, selecting the checkpoint with the best validation force MAE for testing. An exponential moving average (EMA) learning scheduler with a decay factor of 0.99 was used, along with gradient clipping set to 10. The number of MPtraj samples used in the multi-head fine-tuning algorithm was fixed at 400, with a corresponding weight of 1.

\subsection{Supplementary Numerical Results}
\label{sec:apd:numer}

We focus on the mechanical properties of Mo, Ni, and Ge, as shown in Section 3.1. Similar to Table~\ref{tab:metals}, Table~\ref{tab:metals-remain} highlights large discrepancies in the original foundation models, with relative errors ranging from 20\% to 80\%, likely due to limited emphasis on stress during training. Fine-tuned and scratch-trained models show significantly improved accuracy, with lattice parameters within 0.1–2.0\% and elastic constants within 10\% of DFT values. However, fine-tuned models often struggle with vacancy formation energies and migration barriers in BCC and FCC systems, where ACE models trained from scratch perform better.

\begin{table}
\centering
\resizebox{\textwidth}{!}{
\begin{tabular}{cccccccc}
  & DFT & \ml & Fine-tuning-0b & \mlo & Fine-tuning-0 & MACE-scratch & ACE-scratch \\
\hline
  &  &  & & \textbf{Mo} & & & \\
\hline
$a_0$  &  3.168 &  3.177 (+0.3\%) &           3.16 (-0.3\%) &   3.170 (+0.1\%) &  \textbf{3.168 (+0.0\%)} &          3.156 (-0.4\%) &  \textbf{3.168 (+0.0\%)} \\
$C_{11}$ &    472 &   290 (-38.6\%) &            483 (+2.3\%) &   343 (-27.3\%) &             498 (+5.5\%) &           519 (+10.0\%) &    \textbf{472 (+0.0\%)} \\
$C_{12}$ &    158 &   230 (+45.6\%) &           182 (+15.2\%) &   234 (+48.1\%) &             147 (-7.0\%) &            169 (+7.0\%) &    \textbf{159 (+0.6\%)} \\
$C_{44}$ &    106 &    33 (-68.9\%) &             97 (-8.5\%) &    38 (-64.2\%) &             113 (+6.6\%) &            108 (+1.9\%) &    \textbf{107 (+0.9\%)} \\
$B$   &    263 &    250 (-4.9\%) &            285 (+8.4\%) &    270 (+2.7\%) &             264 (+0.4\%) &            286 (+8.7\%) &    \textbf{263 (+0.0\%)} \\
$E_v$  &    2.7 &   1.8 (-33.3\%) &  \textbf{2.65 (-1.9\%)} &  2.23 (-17.4\%) &            2.55 (-5.6\%) &          2.33 (-13.7\%) &           2.32 (-14.1\%) \\
$E_m$  &   1.22 &  0.35 (-71.3\%) &          1.71 (+40.2\%) &  0.91 (-25.4\%) &           1.08 (-11.5\%) &          1.81 (+48.4\%) &  \textbf{1.35 (+10.7\%)} \\
$E_a$ &   3.92 &  2.15 (-45.2\%) &          4.37 (+11.5\%) &  3.14 (-19.9\%) &            3.63 (-7.4\%) &  \textbf{4.14 (+5.6\%)} &            3.67 (-6.4\%) \\
\hline \\

  &  &  &  & \textbf{Ni} & & &  \\
\hline
$a_0$  &  3.508 &  \textbf{3.507 (-0.03\%)} &  \textbf{3.507 (-0.03\%)} &   3.510 (+0.06\%) &           3.510 (+0.06\%) &  3.518 (+0.29\%) &          3.521 (+0.37\%) \\
$C_{11}$ &    276 &            246 (-10.87\%) &             258 (-6.52\%) &    279 (+1.09\%) &   \textbf{278 (+0.72\%)} &    270 (-2.17\%) &            273 (-1.09\%) \\
$C_{12}$ &    159 &            190 (+19.50\%) &            183 (+15.09\%) &   178 (+11.95\%) &   \textbf{172 (+8.18\%)} &   175 (+10.06\%) &           179 (+12.58\%) \\
$C_{44}$ &    132 &            112 (-15.15\%) &             120 (-9.09\%) &    98 (-25.76\%) &           100 (-24.24\%) &   110 (-16.67\%) &   \textbf{123 (-6.82\%)} \\
$B$   &    198 &             209 (+5.56\%) &    \textbf{206 (+4.04\%)} &    211 (+6.57\%) &            209 (+5.56\%) &    207 (+4.55\%) &            211 (+6.57\%) \\
$E_v$  &   1.49 &            1.55 (+4.03\%) &           1.67 (+12.08\%) &  1.17 (-21.48\%) &          1.34 (-10.07\%) &  1.65 (+10.74\%) &  \textbf{1.44 (-3.36\%)} \\
$E_m$  &   1.12 &            1.03 (-8.04\%) &            1.04 (-7.14\%) &  0.82 (-26.79\%) &  \textbf{1.14 (+1.79\%)} &   1.02 (-8.93\%) &   \textbf{1.10 (-1.79\%)} \\
$E_a$ &   2.61 &   \textbf{2.58 (-1.15\%)} &            2.71 (+3.83\%) &  1.99 (-23.75\%) &           2.48 (-4.98\%) &   2.67 (+2.30\%) &           2.54 (-2.68\%) \\
\hline \\

 & &  &  &  \textbf{Ge} &  &  &  \\
\hline
$a_0$  &  5.763 &  5.746 (-0.3\%) &           5.75 (-0.2\%) &        5.773 (+0.2\%) &  \textbf{5.756 (-0.1\%)} &          5.774 (+0.2\%) &  \textbf{5.766 (+0.1\%)} \\
$C_{11}$ &    116 &    61 (-47.4\%) &            77 (-33.6\%) &          59 (-49.1\%) &            100 (-13.8\%) &            78 (-32.8\%) &   \textbf{103 (-11.2\%)} \\
$C_{12}$ &     48 &    37 (-22.9\%) &             44 (-8.3\%) &  \textbf{47 (-2.1\%)} &             75 (+56.2\%) &             46 (-4.2\%) &             80 (+66.7\%) \\
$C_{44}$ &     58 &    15 (-74.1\%) &            38 (-34.5\%) &           3 (-94.8\%) &             50 (-13.8\%) &            32 (-44.8\%) &    \textbf{52 (-10.3\%)} \\
$B$   &     71 &    45 (-36.6\%) &   \textbf{59 (-16.9\%)} &          51 (-28.2\%) &    \textbf{83 (+16.9\%)} &            58 (-18.3\%) &             88 (+23.9\%) \\
$E_v$  &   2.19 &  0.57 (-74.0\%) &  \textbf{2.06 (-5.9\%)} &        0.08 (-96.3\%) &           1.55 (-29.2\%) &           2.39 (+9.1\%) &           3.29 (+50.2\%) \\
$E_m$  &   0.19 &  0.02 (-89.5\%) &          0.11 (-42.1\%) &        0.02 (-89.5\%) &  \textbf{0.15 (-21.1\%)} &           0.10 (-47.4\%) &          0.51 (+168.4\%) \\
$E_a$ &   2.38 &  0.59 (-75.2\%) &           2.17 (-8.8\%) &         0.10 (-95.8\%) &            1.70 (-28.6\%) &  \textbf{2.49 (+4.6\%)} &            3.8 (+59.7\%) \\
\hline
\end{tabular}
}
\caption{Comparison of predicted material properties and percentage errors across methods for Mo, Ni, and Ge.}
\label{tab:metals-remain} 
\end{table}

\begin{figure}
\centering
\includegraphics[height=7cm]{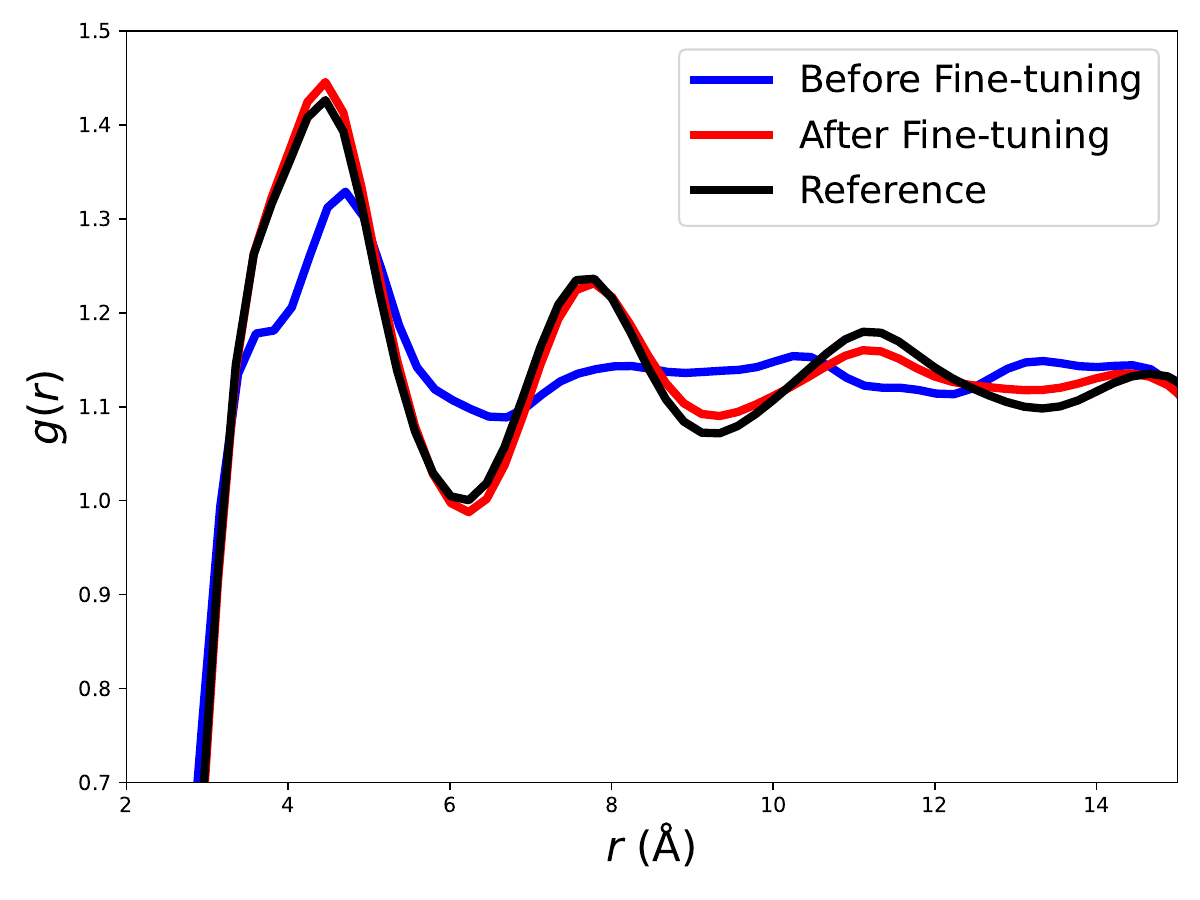}
\caption{Radial distribution function (RDF) of molten LiCl at 900 K, comparing results from \ml~and its fine-tuned MD simulation.}
\label{fig:LiCl-rdf}
\end{figure}

Figure~\ref{fig:LiCl-rdf} shows the radial distribution function (RDF) of molten LiCl at 900 K, comparing the \ml~model before (blue solid line) and after fine-tuning (red solid line). Before fine-tuning, the RDF deviates significantly, with a broader, less pronounced first peak and poorly resolved oscillations at larger distances, indicating inaccuracies in atomic arrangement. After fine-tuning, the RDF improves notably, with a sharper first peak around $r \approx 4 \, \text{\AA}$ and well-defined oscillations, closely matching expected structural properties~\cite{sivaraman2021automated}.

\bibliographystyle{elsarticle-num} 
\bibliography{ft.bib}

\begin{thebibliography}{10}
\expandafter\ifx\csname url\endcsname\relax
  \def\url#1{\texttt{#1}}\fi
\expandafter\ifx\csname urlprefix\endcsname\relax\def\urlprefix{URL }\fi
\expandafter\ifx\csname href\endcsname\relax
  \def\href#1#2{#2} \def\path#1{#1}\fi

\bibitem{behler2016perspective}
J.~Behler, Perspective: Machine learning potentials for atomistic simulations,
  The Journal of chemical physics 145~(17) (2016).

\bibitem{vamathevan2019applications}
J.~Vamathevan, D.~Clark, P.~Czodrowski, I.~Dunham, E.~Ferran, G.~Lee, B.~Li,
  A.~Madabhushi, P.~Shah, M.~Spitzer, et~al., Applications of machine learning
  in drug discovery and development, Nature reviews Drug discovery 18~(6)
  (2019) 463--477.

\bibitem{raccuglia2016machine}
P.~Raccuglia, K.~C. Elbert, P.~D. Adler, C.~Falk, M.~B. Wenny, A.~Mollo,
  M.~Zeller, S.~A. Friedler, J.~Schrier, A.~J. Norquist,
  Machine-learning-assisted materials discovery using failed experiments,
  Nature 533~(7601) (2016) 73--76.

\bibitem{sholl2022density}
D.~S. Sholl, J.~A. Steckel, Density functional theory: a practical
  introduction, John Wiley \& Sons, 2022.

\bibitem{koch2015chemist}
W.~Koch, M.~C. Holthausen, A chemist's guide to density functional theory, John
  Wiley \& Sons, 2015.

\bibitem{ponder2003force}
J.~W. Ponder, D.~A. Case, Force fields for protein simulations, Advances in
  protein chemistry 66 (2003) 27--85.

\bibitem{altona2005empirical}
C.~Altona, D.~H. Faber, Empirical force field calculations: A tool in
  structural organic chemistry, in: Dynamic Chemistry, Springer, 2005, pp.
  1--38.

\bibitem{lee2000second}
B.-J. Lee, M.~I. Baskes, Second nearest-neighbor modified embedded-atom-method
  potential, Physical Review B 62~(13) (2000) 8564.

\bibitem{behler2007generalized}
J.~Behler, M.~Parrinello, Generalized neural-network representation of
  high-dimensional potential-energy surfaces, Physical review letters 98~(14)
  (2007) 146401.

\bibitem{bartok2010gaussian}
A.~P. Bart{\'o}k, M.~C. Payne, R.~Kondor, G.~Cs{\'a}nyi, Gaussian approximation
  potentials: The accuracy of quantum mechanics, without the electrons,
  Physical review letters 104~(13) (2010) 136403.

\bibitem{witt2023acepotentials}
W.~C. Witt, C.~van~der Oord, E.~Gel{\v{z}}inyt{\.e}, T.~J{\"a}rvinen, A.~Ross,
  J.~P. Darby, C.~H. Ho, W.~J. Baldwin, M.~Sachs, J.~Kermode, et~al.,
  Acepotentials. jl: A julia implementation of the atomic cluster expansion,
  The Journal of Chemical Physics 159~(16) (2023).

\bibitem{batatia2022mace}
I.~Batatia, D.~P. Kovacs, G.~Simm, C.~Ortner, G.~Cs{\'a}nyi, Mace: Higher order
  equivariant message passing neural networks for fast and accurate force
  fields, Advances in Neural Information Processing Systems 35 (2022)
  11423--11436.

\bibitem{shapeev2016moment}
A.~V. Shapeev, Moment tensor potentials: A class of systematically improvable
  interatomic potentials, Multiscale Modeling \& Simulation 14~(3) (2016)
  1153--1173.

\bibitem{DrautzACE}
R.~Drautz, Atomic cluster expansion for accurate and transferable interatomic
  potentials, Phys. Rev. B 99 (2019) 014104.

\bibitem{ACECompleteness}
G.~Dusson, M.~Bachmayr, G.~Cs{\'a}nyi, R.~Drautz, S.~Etter, C.~van~der Oord,
  C.~Ortner, Atomic cluster expansion: Completeness, efficiency and stability,
  J. Comput. Phys. 454 (2022) 110946.

\bibitem{luo2024enabling}
S.~Luo, T.~Chen, A.~S. Krishnapriyan, Enabling efficient equivariant operations
  in the fourier basis via gaunt tensor products, arXiv preprint
  arXiv:2401.10216 (2024).

\bibitem{darby2023tensor}
J.~P. Darby, D.~P. Kov{\'a}cs, I.~Batatia, M.~A. Caro, G.~L. Hart, C.~Ortner,
  G.~Cs{\'a}nyi, Tensor-reduced atomic density representations, Physical Review
  Letters 131~(2) (2023) 028001.

\bibitem{gilmer2017neural}
J.~Gilmer, S.~S. Schoenholz, P.~F. Riley, O.~Vinyals, G.~E. Dahl, Neural
  message passing for quantum chemistry, in: International conference on
  machine learning, PMLR, 2017, pp. 1263--1272.

\bibitem{gilmer2020message}
J.~Gilmer, S.~S. Schoenholz, P.~F. Riley, O.~Vinyals, G.~E. Dahl, Message
  passing neural networks, in: Machine learning meets quantum physics,
  Springer, 2020, pp. 199--214.

\bibitem{thirunavukarasu2023large}
A.~J. Thirunavukarasu, D.~S.~J. Ting, K.~Elangovan, L.~Gutierrez, T.~F. Tan,
  D.~S.~W. Ting, Large language models in medicine, Nature medicine 29~(8)
  (2023) 1930--1940.

\bibitem{kasneci2023chatgpt}
E.~Kasneci, K.~Se{\ss}ler, S.~K{\"u}chemann, M.~Bannert, D.~Dementieva,
  F.~Fischer, U.~Gasser, G.~Groh, S.~G{\"u}nnemann, E.~H{\"u}llermeier, et~al.,
  Chatgpt for good? on opportunities and challenges of large language models
  for education, Learning and individual differences 103 (2023) 102274.

\bibitem{chanussot2021open}
L.~Chanussot, A.~Das, S.~Goyal, T.~Lavril, M.~Shuaibi, M.~Riviere, K.~Tran,
  J.~Heras-Domingo, C.~Ho, W.~Hu, et~al., Open catalyst 2020 (oc20) dataset and
  community challenges, Acs Catalysis 11~(10) (2021) 6059--6072.

\bibitem{bowman2022md17}
J.~M. Bowman, C.~Qu, R.~Conte, A.~Nandi, P.~L. Houston, Q.~Yu, The md17
  datasets from the perspective of datasets for gas-phase “small” molecule
  potentials, The Journal of Chemical Physics 156~(24) (2022).

\bibitem{jain2013commentary}
A.~Jain, S.~P. Ong, G.~Hautier, W.~Chen, W.~D. Richards, S.~Dacek, S.~Cholia,
  D.~Gunter, D.~Skinner, G.~Ceder, et~al., Commentary: The materials project: A
  materials genome approach to accelerating materials innovation, APL materials
  1~(1) (2013).

\bibitem{bommasani2021opportunities}
R.~Bommasani, D.~A. Hudson, E.~Adeli, R.~Altman, S.~Arora, S.~von Arx, M.~S.
  Bernstein, J.~Bohg, A.~Bosselut, E.~Brunskill, et~al., On the opportunities
  and risks of foundation models, arXiv preprint arXiv:2108.07258 (2021).

\bibitem{batatia2023foundation}
I.~Batatia, P.~Benner, Y.~Chiang, A.~M. Elena, D.~P. Kov{\'a}cs, J.~Riebesell,
  X.~R. Advincula, M.~Asta, W.~J. Baldwin, N.~Bernstein, et~al., A foundation
  model for atomistic materials chemistry, arXiv preprint arXiv:2401.00096
  (2023).

\bibitem{deng2023chgnet}
B.~Deng, P.~Zhong, K.~Jun, J.~Riebesell, K.~Han, C.~J. Bartel, G.~Ceder, Chgnet
  as a pretrained universal neural network potential for charge-informed
  atomistic modelling, Nature Machine Intelligence 5~(9) (2023) 1031--1041.

\bibitem{merchant2023scaling}
A.~Merchant, S.~Batzner, S.~S. Schoenholz, M.~Aykol, G.~Cheon, E.~D. Cubuk,
  Scaling deep learning for materials discovery, Nature 624~(7990) (2023)
  80--85.

\bibitem{zhang2023dpa}
D.~Zhang, X.~Liu, X.~Zhang, C.~Zhang, C.~Cai, H.~Bi, Y.~Du, X.~Qin, J.~Huang,
  B.~Li, et~al., Dpa-2: Towards a universal large atomic model for molecular
  and material simulation, arXiv preprint arXiv:2312.15492 (2023).

\bibitem{choudhary2023unified}
K.~Choudhary, B.~DeCost, L.~Major, K.~Butler, J.~Thiyagalingam, F.~Tavazza,
  Unified graph neural network force-field for the periodic table: solid state
  applications, Digital Discovery 2~(2) (2023) 346--355.

\bibitem{chen2022universal}
C.~Chen, S.~P. Ong, A universal graph deep learning interatomic potential for
  the periodic table, Nature Computational Science 2~(11) (2022) 718--728.

\bibitem{li2024extendable}
Z.~Li, T.~Wen, Y.~Zhang, X.~Liu, C.~Zhang, A.~Pattamatta, X.~Gong, B.~Ye,
  H.~Wang, L.~Zhang, et~al., An extendable cloud-native alloy property
  explorer, arXiv preprint arXiv:2404.17330 (2024).

\bibitem{focassio2024performance}
B.~Focassio, L.~P. M.~Freitas, G.~R. Schleder, Performance assessment of
  universal machine learning interatomic potentials: Challenges and directions
  for materials’ surfaces, ACS Applied Materials \& Interfaces (2024).

\bibitem{deng2024overcoming}
B.~Deng, Y.~Choi, P.~Zhong, J.~Riebesell, S.~Anand, Z.~Li, K.~Jun, K.~A.
  Persson, G.~Ceder, Overcoming systematic softening in universal machine
  learning interatomic potentials by fine-tuning, arXiv preprint
  arXiv:2405.07105 (2024).

\bibitem{alavi2024towards}
S.~F. Alavi, Y.~Chen, Y.-F. Hou, F.~Ge, P.~Zheng, P.~O. Dral, Towards accurate
  and efficient anharmonic vibrational frequencies with the universal
  interatomic potential ani-1ccx-gelu and its fine-tuning (2024).

\bibitem{yu2024systematic}
H.~Yu, M.~Giantomassi, G.~Materzanini, J.~Wang, G.-M. Rignanese, Systematic
  assessment of various universal machine-learning interatomic potentials,
  Materials Genome Engineering Advances 2~(3) (2024) e58.

\bibitem{pyzer2025foundation}
E.~O. Pyzer-Knapp, M.~Manica, P.~Staar, L.~Morin, P.~Ruch, T.~Laino, J.~R.
  Smith, A.~Curioni, Foundation models for materials discovery--current state
  and future directions, npj Computational Materials 11~(1) (2025) 61.

\bibitem{shuang2025universal}
F.~Shuang, Z.~Wei, K.~Liu, W.~Gao, P.~Dey, Universal machine learning
  interatomic potentials poised to supplant dft in modeling general defects in
  metals and random alloys, arXiv preprint arXiv:2502.03578 (2025).

\bibitem{du2025universal}
H.~Du, J.~Hui, L.~Zhang, H.~Wang, Universal machine learning interatomic
  potentials are ready for solid ion conductors, arXiv preprint
  arXiv:2502.09970 (2025).

\bibitem{lee2025accelerating}
H.~Lee, V.~I. Hegde, C.~Wolverton, Y.~Xia, Accelerating high-throughput phonon
  calculations via machine learning universal potentials, Materials Today
  Physics (2025) 101688.

\bibitem{niblett2024transferability}
S.~P. Niblett, P.~Kourtis, I.-B. Magd{\u{a}}u, C.~P. Grey, G.~Cs{\'a}nyi,
  Transferability of datasets between machine-learning interaction potentials,
  arXiv preprint arXiv:2409.05590 (2024).

\bibitem{casillas2024evaluating}
L.~Casillas-Trujillo, A.~S. Parackal, R.~Armiento, B.~Alling, Evaluating and
  improving the predictive accuracy of mixing enthalpies and volumes in
  disordered alloys from universal pretrained machine learning potentials,
  Physical Review Materials 8~(11) (2024) 113803.

\bibitem{wang2024impact}
L.~Wang, T.~He, B.~Ouyang, The impact of domain knowledge on universal machine
  learning models, in: arXiv, 2024.

\bibitem{RBM_Jin}
S.~Jin, L.~Li, J.-G. Liu, Random batch methods (rbm) for interacting particle
  systems, Journal of Computational Physics 400 (2020) 108877.

\bibitem{RBE}
S.~Jin, L.~Li, Z.~Xu, Y.~Zhao, A random batch ewald method for particle systems
  with coulomb interactions, SIAM Journal on Scientific Computing 43~(4) (2021)
  B937--B960.

\bibitem{RBE_2}
J.~Liang, P.~Tan, Y.~Zhao, L.~Li, S.~Jin, L.~Hong, Z.~Xu, Superscalability of
  the random batch ewald method, The Journal of Chemical Physics 156~(1)
  (2022).

\bibitem{IRBE}
J.~Liang, Z.~Xu, Y.~Zhao, Improved random batch ewald method in molecular
  dynamics simulations, The Journal of Physical Chemistry A 126~(22) (2022)
  3583--3593.

\bibitem{gao2024rbmd}
W.~Gao, T.~Zhao, Y.~Guo, J.~Liang, H.~Liu, M.~Luo, Z.~Luo, W.~Qin, Y.~Wang,
  Q.~Zhou, et~al., Rbmd: A molecular dynamics package enabling to simulate 10
  million all-atom particles in a single graphics processing unit, arXiv
  preprint arXiv:2407.09315 (2024).

\bibitem{ho2024atomic}
C.~H. Ho, T.~S. Gutleb, C.~Ortner, Atomic cluster expansion without
  self-interaction, arXiv preprint arXiv:2401.01550 (2024).

\bibitem{ACEHam}
L.~Zhang, B.~Onat, G.~Dusson, A.~McSloy, G.~Anand, R.~J. Maurer, C.~Ortner,
  J.~R. Kermode, Equivariant analytical mapping of first principles
  {Hamiltonians} to accurate and transferable materials models, npj Comput.
  Mater. 8~(1) (2022) 158.

\bibitem{wang2024theoretical}
Y.~Wang, S.~Patel, C.~Ortner, A theoretical case study of the generalization of
  machine-learned potentials, Computer Methods in Applied Mechanics and
  Engineering 422 (2024) 116831.

\bibitem{torabi2024surrogate}
T.~Torabi, Y.~Wang, C.~Ortner, Surrogate models for vibrational entropy based
  on a spatial decomposition, arXiv preprint arXiv:2402.12744 (2024).

\bibitem{wang2025many}
Y.~Wang, G.~Csanyi, C.~Ortner, Many-body coarse-grained molecular dynamics with
  the atomic cluster expansion, arXiv preprint arXiv:2502.04661 (2025).

\bibitem{vignac2020building}
C.~Vignac, A.~Loukas, P.~Frossard, Building powerful and equivariant graph
  neural networks with structural message-passing, Advances in neural
  information processing systems 33 (2020) 14143--14155.

\bibitem{maron2018invariant}
H.~Maron, H.~Ben-Hamu, N.~Shamir, Y.~Lipman, Invariant and equivariant graph
  networks, arXiv preprint arXiv:1812.09902 (2018).

\bibitem{tajbakhsh2016convolutional}
N.~Tajbakhsh, J.~Y. Shin, S.~R. Gurudu, R.~T. Hurst, C.~B. Kendall, M.~B.
  Gotway, J.~Liang, Convolutional neural networks for medical image analysis:
  Full training or fine tuning?, IEEE transactions on medical imaging 35~(5)
  (2016) 1299--1312.

\bibitem{jordan2015machine}
M.~I. Jordan, T.~M. Mitchell, Machine learning: Trends, perspectives, and
  prospects, Science 349~(6245) (2015) 255--260.

\bibitem{zhang2022fine}
H.~Zhang, G.~Li, J.~Li, Z.~Zhang, Y.~Zhu, Z.~Jin, Fine-tuning pre-trained
  language models effectively by optimizing subnetworks adaptively, Advances in
  Neural Information Processing Systems 35 (2022) 21442--21454.

\bibitem{devlin2019bert}
J.~Devlin, M.-W. Chang, K.~Lee, K.~Toutanova, Bert: Pre-training of deep
  bidirectional transformers for language understanding, in: Proceedings of the
  2019 Conference of the North American Chapter of the Association for
  Computational Linguistics: Human Language Technologies, 2019, pp. 4171--4186.

\bibitem{lee2020biobert}
J.~Lee, W.~Yoon, S.~Kim, D.~Kim, S.~Kim, C.~H. So, J.~Kang, Biobert: a
  pre-trained biomedical language model for biomedical text mining,
  Bioinformatics 36~(4) (2020) 1234--1240.

\bibitem{hyperactive2022}
C.~Oord, M.~Sachs, D.~Kovacs, C.~Ortner, G.~Csányi, Hyperactive learning for
  data-driven interatomic potentials, npj Computational Materials 9 (09 2023).

\bibitem{voita2019analyzing}
E.~Voita, D.~Talbot, F.~Moiseev, R.~Sennrich, I.~Titov, Analyzing multi-head
  self-attention: Specialized heads do the heavy lifting, the rest can be
  pruned, arXiv preprint arXiv:1905.09418 (2019).

\bibitem{kim2024hydra}
S.~Kim, H.~Yang, Y.~Kim, Y.~Hong, E.~Park, Hydra: Multi-head low-rank
  adaptation for parameter efficient fine-tuning, Neural Networks (2024)
  106414.

\bibitem{Zuo2020}
Y.~Zuo, C.~Chen, X.~Li, Z.~Deng, Y.~Chen, J.~Behler, G.~Csányi, A.~V. Shapeev,
  A.~P. Thompson, M.~A. Wood, S.~P. Ong, Performance and cost assessment of
  machine learning interatomic potentials, J. Phys. Chem. A 124~(4) (2020)
  731--745, pMID: 31916773.

\bibitem{henkelman2000climbing}
G.~Henkelman, B.~P. Uberuaga, H.~J{\'o}nsson, A climbing image nudged elastic
  band method for finding saddle points and minimum energy paths, The Journal
  of chemical physics 113~(22) (2000) 9901--9904.

\bibitem{bartok2018machine}
A.~P. Bart{\'o}k, J.~Kermode, N.~Bernstein, G.~Cs{\'a}nyi, Machine learning a
  general-purpose interatomic potential for silicon, Physical Review X 8~(4)
  (2018) 041048.

\bibitem{toit2024hyperparameter}
D.~F. Toit, Y.~Zhou, V.~L. Deringer, Hyperparameter optimization for atomic
  cluster expansion potentials, arXiv preprint arXiv:2408.00656 (2024).

\bibitem{zhang2023atomistic}
L.~Zhang, G.~Cs{\'a}nyi, E.~Van Der~Giessen, F.~Maresca, Atomistic fracture in
  bcc iron revealed by active learning of gaussian approximation potential, npj
  Computational Materials 9~(1) (2023) 217.

\bibitem{guenole2020assessment}
J.~Gu{\'e}nol{\'e}, W.~G. N{\"o}hring, A.~Vaid, F.~Houll{\'e}, Z.~Xie,
  A.~Prakash, E.~Bitzek, Assessment and optimization of the fast inertial
  relaxation engine (fire) for energy minimization in atomistic simulations and
  its implementation in lammps, Computational Materials Science 175 (2020)
  109584.

\bibitem{grigorev2024matscipy}
P.~Grigorev, L.~Fr{\'e}rot, F.~Birks, A.~Gola, J.~Golebiowski, J.~Grie{\ss}er,
  J.~L. H{\"o}rmann, A.~Klemenz, G.~Moras, W.~G. N{\"o}hring, et~al., matscipy:
  materials science at the atomic scale with python, Journal of Open Source
  Software 9~(93) (2024).

\bibitem{byggmastar2019machine}
J.~Byggm{\"a}star, A.~Hamedani, K.~Nordlund, F.~Djurabekova, Machine-learning
  interatomic potential for radiation damage and defects in tungsten, Physical
  Review B 100~(14) (2019) 144105.

\bibitem{miracle2017critical}
D.~B. Miracle, O.~N. Senkov, A critical review of high entropy alloys and
  related concepts, Acta materialia 122 (2017) 448--511.

\bibitem{george2019high}
E.~P. George, D.~Raabe, R.~O. Ritchie, High-entropy alloys, Nature reviews
  materials 4~(8) (2019) 515--534.

\bibitem{stukowski2009visualization}
A.~Stukowski, Visualization and analysis of atomistic simulation data with
  ovito--the open visualization tool, Modelling and simulation in materials
  science and engineering 18~(1) (2009) 015012.

\bibitem{gou2021knowledge}
J.~Gou, B.~Yu, S.~J. Maybank, D.~Tao, Knowledge distillation: A survey,
  International Journal of Computer Vision 129~(6) (2021) 1789--1819.

\bibitem{alkhulaifi2021knowledge}
A.~Alkhulaifi, F.~Alsahli, I.~Ahmad, Knowledge distillation in deep learning
  and its applications, PeerJ Computer Science 7 (2021) e474.

\bibitem{sivaraman2021automated}
G.~Sivaraman, J.~Guo, L.~Ward, N.~Hoyt, M.~Williamson, I.~Foster, C.~Benmore,
  N.~Jackson, Automated development of molten salt machine learning potentials:
  application to licl, The Journal of Physical Chemistry Letters 12~(17) (2021)
  4278--4285.

\bibitem{liu2024dora}
S.-Y. Liu, C.-Y. Wang, H.~Yin, P.~Molchanov, Y.-C.~F. Wang, K.-T. Cheng, M.-H.
  Chen, Dora: Weight-decomposed low-rank adaptation, arXiv preprint
  arXiv:2402.09353 (2024).

\bibitem{allen2024learning}
A.~E. Allen, N.~Lubbers, S.~Matin, J.~Smith, R.~Messerly, S.~Tretiak,
  K.~Barros, Learning together: Towards foundation models for machine learning
  interatomic potentials with meta-learning, npj computational materials 10~(1)
  (2024) 154.

\bibitem{matin2025teacher}
S.~Matin, A.~Allen, E.~Shinkle, A.~Pachalieva, G.~T. Craven, B.~Nebgen,
  J.~Smith, R.~Messerly, Y.~W. Li, S.~Tretiak, et~al., Teacher-student training
  improves accuracy and efficiency of machine learning inter-atomic potentials,
  arXiv preprint arXiv:2502.05379 (2025).

\bibitem{chen2022qm}
H.~Chen, C.~Ortner, Y.~Wang, Qm/mm methods for crystalline defects. part 3:
  Machine-learned mm models, Multiscale Modeling \& Simulation 20~(4) (2022)
  1490--1518.

\bibitem{wang2021posteriori}
Y.~Wang, H.~Chen, M.~Liao, C.~Ortner, H.~Wang, L.~Zhang, A posteriori error
  estimates for adaptive qm/mm coupling methods, SIAM Journal on Scientific
  Computing 43~(4) (2021) A2785--A2808.

\bibitem{fu2025meshac}
K.~Fu, M.~Liao, Y.~Wang, J.~Chen, L.~Zhang, Meshac: A 3d mesh generation and
  adaptation package for multiscale coupling methods, Computer Physics
  Communications (2025) 109523.

\bibitem{fu2023adaptive}
K.~Fu, M.~Liao, Y.~Wang, J.~Chen, L.~Zhang, Adaptive multigrid strategy for
  geometry optimization of large-scale three dimensional molecular mechanics,
  Journal of Computational Physics 485 (2023) 112113.

\bibitem{wang2025posteriori}
Y.~Wang, A posteriori analysis and adaptive algorithms for blended type
  atomistic-to-continuum coupling with higher-order finite elements, Computer
  Physics Communications (2025) 109533.

\bibitem{siddiqui2024machine}
A.~Siddiqui, N.~D. Hine, Machine-learned interatomic potentials for transition
  metal dichalcogenide mo1- x w x s2- 2 y se2 y alloys, npj Computational
  Materials 10~(1) (2024) 169.

\bibitem{vandenhaute2023machine}
S.~Vandenhaute, M.~Cools-Ceuppens, S.~DeKeyser, T.~Verstraelen,
  V.~Van~Speybroeck, Machine learning potentials for metal-organic frameworks
  using an incremental learning approach, npj Computational Materials 9~(1)
  (2023) 1--8.

\bibitem{harper2020ab}
A.~F. Harper, M.~L. Evans, J.~P. Darby, B.~Karasulu, C.~P. Ko{\c{c}}er, J.~R.
  Nelson, A.~J. Morris, Ab initio structure prediction methods for battery
  materials: A review of recent computational efforts to predict the atomic
  level structure and bonding in materials for rechargeable batteries, Johnson
  Matthey Technology Review 64~(2) (2020) 103--118.

\bibitem{wang2022anisotropic}
J.~Wang, C.~Jiang, W.~Li, X.~Xiao, Anisotropic low-dimensional materials for
  polarization-sensitive photodetectors: from materials to devices, Advanced
  Optical Materials 10~(6) (2022) 2102436.

\bibitem{paszke2019pytorch}
A.~Paszke, S.~Gross, F.~Massa, A.~Lerer, J.~Bradbury, G.~Chanan, T.~Killeen,
  Z.~Lin, N.~Gimelshein, L.~Antiga, et~al., Pytorch: An imperative style,
  high-performance deep learning library, Advances in neural information
  processing systems 32 (2019).

\bibitem{geiger2022e3nn}
M.~Geiger, T.~Smidt, e3nn: Euclidean neural networks, arXiv preprint
  arXiv:2207.09453 (2022).

\bibitem{kingma2014adam}
D.~P. Kingma, J.~Ba, Adam: A method for stochastic optimization, arXiv preprint
  arXiv:1412.6980 (2014).

\end{thebibliography}





\end{document}